\pdfoutput=1
\documentclass[a4paper,12pt]{article}
\usepackage{jcappub}
\usepackage[utf8]{inputenc}

\usepackage[table]{xcolor}
\usepackage{amsmath,amssymb}
\usepackage{graphicx}
\usepackage{multicol}
\usepackage{hyperref}
\usepackage{physics}
\usepackage[super]{nth}
\usepackage{slashed}
\usepackage{orcidlink}

\newcommand{\GeV}{\; \mathrm{GeV}}
\newcommand{\TeV}{\; \mathrm{TeV}}
\newcommand{\beq}{\begin{equation}}
\newcommand{\eeq}{\end{equation}}
\newcommand{\bea}{\begin{eqnarray}}
\newcommand{\eea}{\end{eqnarray}}
\newcommand{\mgr}{m_{3/2}}
\newcommand{\grav}{\widetilde{G} }

\newcommand{\mplanck}{\ensuremath{M_{\text{P}}}}
\newcommand{\mpl}{\ensuremath{M_{\text{P}}}}
\newcommand{\suthree}{\ensuremath{\text{SU}(3)_{\text{c}}}}
\newcommand{\sutwo}{\ensuremath{\text{SU}(2)_{\text{L}}}}
\newcommand{\uone}{\ensuremath{\text{U}(1)_{\text{Y}}}}
\newcommand{\sg}{{\tilde g}}
\newcommand{\sq}{{\tilde q}}
\newcommand{\Gr}{{\grav}}
\newcommand{\Grb}{{\overline\grav}}
\newcommand{\nn}{\nonumber \\}

\newcommand\htlsim{\mathrel{\stackrel{\makebox[0pt]{\mbox{\normalfont\tiny HTL}}}{\normalfont \sim}}}

\DeclareMathOperator{\Rea}{Re}

\def\a{\alpha}

\def\d{\delta}
\def\e{\epsilon}
\def\f{\phi}
\def\g{\gamma}

\def\l{\lambda}
\def\m{\mu}
\def\n{\nu}
\def\o{\omega}

\def\r{\rho}
\def\s{\sigma}

\def\D{\Delta}

\def\G{\Gamma}

\def\P{\Pi}

\def\S{\Sigma}

\makeatletter

\def\slash{\@ifnextchar[{\fmsl@sh}{\fmsl@sh[0mu]}}
\def\fmsl@sh[#1]#2{%
  \mathchoice
    {\@fmsl@sh\displaystyle{#1}{#2}}%
    {\@fmsl@sh\textstyle{#1}{#2}}%
    {\@fmsl@sh\scriptstyle{#1}{#2}}%
    {\@fmsl@sh\scriptscriptstyle{#1}{#2}}}
\def\@fmsl@sh#1#2#3{\m@th\ooalign{$\hfil#1\mkern#2/\hfil$\crcr$#1#3$}}
\makeatother
\preprintnumber{FTPI-MINN-24/20}

\title{Gravitino Thermal Production, Dark Matter, and Reheating of the Universe}

\author[\,a]{Helmut~Eberl~\orcidlink{0000-0002-1060-4700}}
\author[\,b]{\!, Ioannis D.~Gialamas~\orcidlink{0000-0002-2957-5276}}
\author[\,c,d]{, Vassilis C.~Spanos~\orcidlink{0000-0001-8676-3655}}

\emailAdd{helmut.eberl@oeaw.ac.at}
\emailAdd{ioannis.gialamas@kbfi.ee}
\emailAdd{vspanos@phys.uoa.gr}

\vspace{5cm}

\affiliation[a]{\it Institut f\"ur Hochenergiephysik der \"Osterreichischen Akademie
der Wissenschaften, \\
\it A--1050 Vienna, Austria}

\affiliation[b]{\it Laboratory of High Energy and Computational Physics, 
National Institute of Chemical Physics and Biophysics, R{\"a}vala pst.~10, Tallinn, 10143, Estonia}

\affiliation[c] {\it  William I. Fine Theoretical Physics Institute, School of
 Physics and Astronomy,\\ University of Minnesota, 116 Church Street S.E., Minneapolis, MN 55455,
 USA} 
\affiliation[d]{\em Section of Nuclear and Particle Physics, Department of Physics,    \\
 National and Kapodistrian University of Athens, GR-15784 Athens, Greece  }

\abstract{We present a full one-loop calculation of the gravitino thermal production rate,  beyond the so-called hard thermal loop approximation, using the corresponding thermal spectral functions in numerical form  on both 
sides of the light cone. This framework requires a full numerical evaluation. We interpret our results within the framework of a general supergravity-based model, remaining agnostic about the specifics of supersymmetry breaking.
In this context, assuming that gravitinos constitute the entirety of the dark matter in the Universe imposes strict constraints on the reheating temperature. 
For example, with a gluino mass at the current LHC limit, a maximum reheating temperature of $T_\mathrm{reh} \simeq 10^9 \GeV$ is compatible with a gravitino mass of $m_{3/2} \simeq 1 \TeV$. Additionally, with a reheating temperature an order of magnitude lower at $T_\mathrm{reh} \simeq 10^8 \GeV$, the common gaugino  mass $M_{1/2}$ can range from $2$ to $4 \TeV$ within the same gravitino mass range. For much higher values of $M_{1/2}$, which are favored by current accelerator and cosmological data in the context of supersymmetric models, such as $M_{1/2} = 10 \TeV$,  and for $m_{3/2} \simeq 1 \TeV$ the reheating temperature compatible  with the gravitino dark matter scenario is $  10^7 \GeV$. If other dark matter particles are considered, the reheating temperature could be much lower. 

}

\begin{document}

\maketitle

\newpage

%%%%%%%%%%%%%%%%%%%%%%%%%%%%%%%%%%%%%%%%%%%%%%
\section{Introduction} \label{sec:intro}

Although  supersymmetry (SUSY) and its local extension, supergravity (SUGRA), lack experimental verification, they 
still remain a promising comprehensive  framework  for new physics beyond the standard model (SM)  of particle physics. 
In the context of these models,   two particles can  play  the role of the dark matter (DM) in the Universe: the
lightest neutralino, an admixture of neutral gauginos and higgsinos, and the gravitino. The latter scenario is 
favoured by the negative direct DM detection results, till now. The neutralino abundance has 
been calculated precisely \cite{Belanger:2004yn, Alguero:2023zol,Ellis:2022emx}, without any important approximation. On the other hand, 
the gravitino's abundance calculation is quite different, as it interacts 
purely gravitationally with the plasma. Therefore, the gravitino is the prototype of a freeze-in  DM particle  and it is produced mainly thermally.

Since,  the dominant production of gravitinos takes place at a quite high temperature, they are assumed to be massless. In this regime 
the so-called non-derivative approach~\cite{Moroi2005}, involving only the spin  $1/2$ goldstino  components 
was used. 
In this effective theory approach, some production amplitudes exhibit infrared (IR) divergences~\cite{Bolz:1998ek,Bolz:2000fu}.
Therefore, they were regularised either using a finite thermal gaugino  mass or an angular cutoff,
in the phase-space  integration~\cite{Bolz:2000xi,Pradler:2006tpx}.
In~\cite{Ellis:1984eq} the basic $2 \to 2$ gravitino production  processes
    had been   classified  and  calculated. 
This calculation was revised in~\cite{Moroi:1993mb, Kawasaki:1994af}.  
Furthermore, in~\cite{Ellis:1995mr} the authors applied the Braaten-Pisarski-Yuan (BPY) method~\cite{Braaten:1989mz,Braaten:1991dd,Braaten:1991gm} for the calculation of the gravitino production rate,
in the HTL limit.
In~\cite{Bolz:1998ek} the IR singularities were regularized as described above using,
finite thermal masses for gauginos. 
Moreover, in~\cite{Bolz:2000fu,Pradler:2006qh}  the BPY method was employed, 
taking in addition  into account the contribution of the spin $3/2$  gravitino components.
Since this contribution is of the order ${M^2_N /  m^2_{3/2}} $, where  
$M_{1,2,3}$, and $m_{3/2}$ are the gauginos  and gravitino mass respectively, their result 
is valid also in the finite gravitino mass regime.

The calculation method developed even more  in~\cite{Rychkov:2007uq}.
The authors   argued that   the basic condition in order to apply
 the BPY prescription, that is   $g \ll 1$, where  $g$ is the gauge coupling constant,
 is not satisfied for every temperature,  especially  if  $g$ is the strong coupling constant. 
 For this reason,    the  full  one-loop thermal gravitino self-energy must be calculated 
   numerically      beyond the hard thermal loop (HTL)   approximation.
The main idea is that the one-loop gravitino selfenergy through the optical theorem  contains 
all the   $1 \to 2$ processes and the majority of the   $2 \to 2$ (refer also to~\cite{Eberl:2013npa,Eberl:2015dia} for three-body gravitino decays in the MSSM). The
$2 \to 2$ processes that are not related to the  selfenergy is the so-called  subtracted part and must be calculated separately. 
 Interestingly enough, this part is free of IR singularities~\cite{Rychkov:2007uq}.   
 
Recently, we have revisited this calculation~\cite{Eberl:2020fml}.   
We  corrected errors of the previous analysis and we presented a new 
numerical result that was differed from that in~\cite{Rychkov:2007uq} by almost $10\%$.
In this paper we provide  all the analytical tools and   results that are required 
for the calculation  of the gravitino selfenergy. Moreover, in our numerical 
calculation we use the full one-loop numerical result for the spectral functions.
This way one can calculate the gravitino selfenergy without any 
numerical approximation in the whole frequency-momentum plane, that 
is both inside and outside of the light cone.
 This more detailed numerical treatment yields an updated result for
 the thermal gravitino production rate, which is approximately 
 40\% lower than that reported in~\cite{Eberl:2020fml}. 
 This reduction diminishes with increasing temperatures, reaching around 
 20\% at $ T \sim 10^8 \GeV$, and disappears entirely at $T \sim 10^{16} \GeV$.

Since the gravitino abundance is related to the maximum reheating temperature, $T_\mathrm{reh}$,
the reduced production rate allows for higher   $T_\mathrm{reh}$ for a given $m_{3/2}$,
as required for  successful thermal leptogenesis~\cite{Barbieri:1999ma,Davidson:2002qv,Giudice:2003jh,Pilaftsis:2003gt,Raidal:2004vt,Antusch:2006gy,Nardi:2006fx,Abada:2006ea}.
The cosmological consequences of this will be discussed in details below.

Depending on the cosmological period, gravitinos may  be produced in various ways.
(a) They can be produced through the  inflaton decays~\cite{Ellis:1982yb,Nanopoulos:1983up,Kallosh:1999jj,Giudice:1999am,Nilles:2001ry,Kawasaki:2006gs,Endo:2006qk,Ellis:2015jpg,Dudas:2017rpa,Kaneta:2019zgw,Kaneta:2023uwi}, (b) thermally, as the Universe cools  down from the  
  $T_\mathrm{reh}$ until now~\cite{Weinberg:1982zq,Ellis:1984eq,Khlopov:1984pf,Moroi:1993mb,Kawasaki:1994af,Moroi:1995fs,Ellis:1995mr,Bolz:1998ek,Bolz:2000fu,Bolz:2000xi,Steffen:2006hw,Pradler:2006qh,Pradler:2006hh,Rychkov:2007uq,Pradler:2006tpx,Ellis:2015jpg, Eberl:2020fml} and (c) through  
 the decays of   unstable particles~\cite{Kawasaki:2008qe,Kawasaki:2017bqm,Cyburt:2006uv,Cyburt:2012kp}  around  the big bang nucleosynthesis (BBN). 
 In the context of the gauge mediated supersymmetry breaking scenario,
 a different production mechanism is used~\cite{Giudice:1998bp,Choi:1999xm,Asaka:2000zh,Jedamzik:2005ir,Fukushima:2013vxa}\footnote{See also~\cite{Leigh:1995jw,Fujisaki1996,Hashimoto1998,Lemoine1999,Maroto2000,Lyth2000a,Takayama2000,Feng2003,Feng2003a,Ellis:2003dn,Kohri2004,Copeland2005,Allahverdi2005,Ellis:2006vu,Khlopov:2004tn,Kawasaki:2006hm,Asaka:2006bv,Cerdeno:2005eu,Buchmuller:2009xv,Cyburt:2009pg,Cyburt:2010vz,Feng2010,Cheung2011,Dalianis:2011ic,Dalianis:2013pya,Badziak:2015dyc,Antusch:2015tha,Ema2016,Arya:2016fnf,Nakayama2014,Garcia:2017tuj,Benakli:2017whb,Co2017,Garcia2018,Dudas:2018npp,Ellis:2019opr,Gu:2020ozv,Kawai:2021hvs,Ahmed:2021dvo,Choi:2021uhy,Afzal:2022vjx,Ahmed:2022rwy,Deshpande:2023zed,Deng:2024rxh} for various works on gravitino production and DM.}.
Recently, an alternative scenario has gained attention, focusing on the so-called ``catastrophic" non-thermal production of slow gravitinos~\cite{Hasegawa:2017hgd,Kolb:2021xfn,Kolb:2021nob,Dudas:2021njv, Terada:2021rtp,Antoniadis:2021jtg}. Additionally, studies on gravitino 
and swampland conjectures have appeared in~\cite{Cribiori:2021gbf,Castellano:2021yye}, 
although the realization of inflation in this scenario appears to be quite challenging.
More comprehensive studies on the production of spin-$3/2$ particles can be found in~~\cite{Garcia2020,Criado:2020jkp,Criado:2021itq,Kaneta:2023uwi,Casagrande:2023fjk,Cirelli:2024ssz}, while constraints on their mass based on the recent $(g-2)_\mu$ measurement are discussed in~\cite{Criado:2021qpd}. 

In our analysis, we will remain agnostic about the specifics 
of SUSY breaking and, consequently, about which particle plays the role of DM.
In SUGRA-based models, there are numerous 
candidates\footnote{Recently, an alternative scenario
has gained attention, even within the 
framework of SUGRA~\cite{Kawasaki:2016pql,Gao:2018pvq,Aldabergenov:2020bpt,Nanopoulos:2020nnh,Wu:2021zta,Spanos:2021hpk,Stamou:2024xkk}:
the production of primordial black holes.
These black holes could potentially serve as dark matter, despite their non-particle origin.} for this role: 
neutral gauginos or specific mixtures of them (such as the neutralino), sneutrinos, axions, axinos, and, last but not least, the gravitino.
In popular and well-studied models within this context, such as the minimal   mSUGRA and the Constrained Minimal Supersymmetric Standard Model (CMSSM), the natural DM particle, beyond the gravitino, is the neutralino.

In the CMSSM, where $m_{3/2}$ is not related to the other soft 
SUSY-breaking masses, the lightest supersymmetric particle 
(LSP) and DM candidate can be either the gravitino or the neutralino.
It is important to note that there will be thermal production of gravitinos in either case.
If the gravitino is the LSP and the DM particle, there are at least two additional production mechanisms for gravitinos, beyond thermal production, that must be considered. These mechanisms will likely necessitate a reduction in the reheating temperature to satisfy the DM constraints.
Gravitinos can be produced during reheating from the decay of the inflaton.
Moreover, if the gravitino ($\grav$) is the LSP the  next-to-lightest supersymmetric particle (NLSP) can be 
the neutralino ($\chi$). Thus, the neutralino is unstable and  will undergo gravitational decay via, for instance,   
$ \chi  \to \tilde{G} \gamma$ with a width~\cite{Eberl:2015dia}
\beq
\Gamma_\chi \sim \frac{m_\chi^3}{\mplanck^2} \, .
\eeq
These late decays, with a timescale of seconds or longer, inject 
electromagnetic energy, and possible hadrons, through three body channels, like
$\chi \to \grav W^- \chi^+$,   during the BBN 
epoch and can therefore affect  its predictions for the abundances of
light elements~\cite{Kohri:2005wn,Cyburt:2012kp}.
Additionally, these decays produce gravitino DM, with
\beq
 \Omega_{3/2} h^2 = \frac{m_{3/2}}{m_\chi} \,   \Omega_{\chi} h^2 \,.
\label{eq:extra_relic1}
\eeq
For these reasons, in the most realistic models, the value  $\Omega_{3/2} h^2 =0.12$  
serves as the upper bound for the thermal DM production, influencing the 
reheating temperature as we will discuss below. 

For the sake of completeness in this discussion, 
it is worth noting that a similar situation 
arises in the neutralino DM scenario, where the gravitino 
can act as the NSLP.
In this case, 
the gravitino  is unstable and  will   decay like~\cite{Ellis:1984er},  
$ \tilde{G}  \to \chi \gamma$ with a width~\cite{Eberl:2013npa}
\beq
\Gamma_{\grav}  \sim \frac{m_{3/2}^3}{\mplanck^2} \, .
\label{eq:chi_decay}
\eeq
Similar to the case of gravitino DM, 
these late decays result in electromagnetic and/or hadronic injections during
BBN, affecting  its predictions~\cite{Kawasaki:2017bqm,Cyburt:2010vz} 
and producing neutralino DM through the 
equation~\eqref{eq:extra_relic1}, only now as
\beq
\Omega_{\chi} h^2 = \frac{m{_\chi}}  {m_{3/2}} \,   \Omega_{3/2} h^2 . 
\label{eq:extra_relic2}
\eeq

In our discussion of the numerical results below, 
we will see that if gravitinos constitute all of the
DM in the Universe, meaning $\Omega_{3/2}h^2 = \Omega_{\rm DM}h^2 \sim 0.12$, 
this imposes strict limits on the reheating temperature. For example, with a gluino mass near the current LHC limit, a reheating temperature of $T_\mathrm{reh} \simeq 10^9 \, \mathrm{GeV}$ is consistent with a gravitino mass of $ m_{3/2} \simeq 1 \, \TeV$. 

For higher $M_{1/2}$ values favored by current accelerator and cosmological data, 
such as $ M_{1/2} = 10 \TeV$ with $ m_{3/2} \simeq 1  \TeV $, the compatible reheating temperature drops to $T_{\rm reh} \simeq 10^7  \GeV$. If we consider other DM candidates like the neutralino, the constraints on the gravitino can be relaxed, allowing for a much lower reheating temperature. Specifically, for $ m_{3/2} \simeq 1  \TeV $
and $ M_{1/2} = 10\TeV $ (or $ M_{1/2} = 20\TeV$), the upper bound on $T_{\rm reh}$ is reduced to $10^4 \, \GeV$  (or  $ 10^3 \GeV$), respectively.

The paper is structured as follows: In section~\ref{gen_fram}, we discuss the general framework of our computation. 
Section~\ref{2to2sub} focuses on calculating the $2\to 2$ amplitudes and the various components of the production rate. Section~\ref{sec:tot_res} presents the overall results, while in section~\ref{sec:abund}, we explore the cosmological implications of our findings, particularly concerning the reheating temperature, which is a crucial factor in calculating other important cosmological quantities, 
such as the baryon asymmetry. In section~\ref{sec:concl}, we present our conclusions, while the extensive appendices provide all the technical details and methodologies used to obtain our results.

\section{General framework}
\label{gen_fram}

Interactions involving gravitinos, $\widetilde{G}$,  are gravitationally suppressed by the inverse of the reduced Planck mass
$\mplanck=(8\pi\, G)^{-1/2}$ (refer to Appendix~\ref{appendix:A} for more details on the gravitino interactions). Consequently, the leading order contributions to gravitino production occur through processes of the form
\beq
a\,b \rightarrow c\, \widetilde{G}\,,
\eeq
where $a$, $b$, and  $c$ can be either three SUSY particles or a combination of one SUSY and two SM particles. Out of ten possible processes, only five need to be explicitly calculated. The remaining five processes can be derived by using crossing symmetries, as we will discuss later. Nevertheless, instead of computing the full amplitudes (${\cal M}$) for the ten possible processes (see table~\ref{table:1} for the corresponding  squared amplitudes for the
$\suthree$ gauge group), i.e.\footnote{The indices $s$, $t$, $u$ denote the diagrams resulting from the exchange of a particle in the respective channel, while the index $x$ stands for the diagram  involving a quartic vertex.}
\beq
\label{eq:full_amp}
| {\cal M}_{X,\rm full} |^2 = |{\cal M}_{X,s} + {\cal M}_{X,t} + {\cal M}_{X,u} + {\cal M}_{X,x} |^2 \, ,
\eeq
with $X=A,B,\cdots ,J$, we will compute the gravitino thermal production rate ($\gamma_{3/2}$) using the following recipe~\cite{Rychkov:2007uq}:
\beq
\gamma_{3/2}  = \gamma_{\rm D} + \gamma_{\rm sub} +\gamma_{\rm top}\,.
\label{eq:gamma32}
\eeq
We will briefly discuss the three contributions and provide more details in the following sections.

\begin{figure}[t!]
 \centering \includegraphics[width=0.35\textwidth]{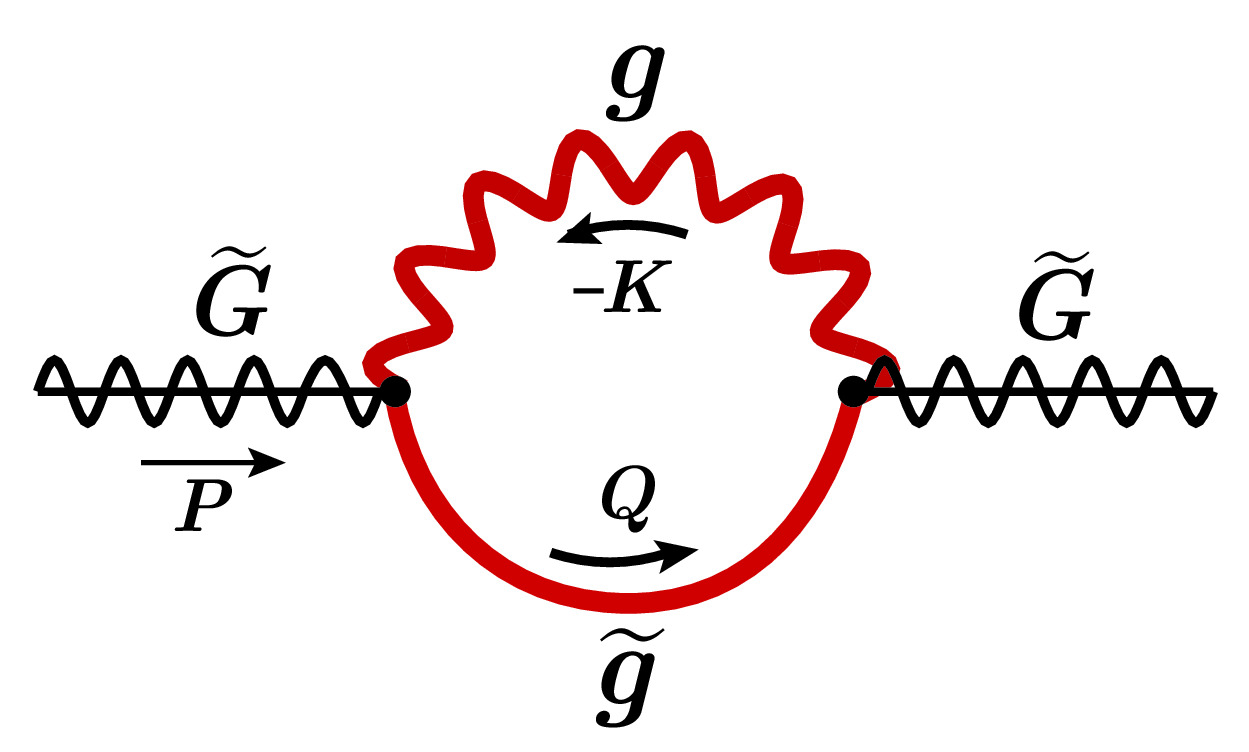}
\caption{The  one-loop  thermally corrected gravitino propagator (D-graph). The 
thick red lines in the graph indicate the gaugino and gauge boson thermally corrected propagators,
as discussed in the text.   }
\label{D_GRAPH}
\end{figure}

\textit{1. The D-graph} ($\gamma_{\rm D}$):  This contribution, illustrated in figure~\ref{D_GRAPH} for the case of the gluon-gluino loop, contains the sum of the squared amplitudes for the $s$, $t$, and $u$ channels,
\beq
|{\cal M}_{X,D}|^2 = |{\cal M}_{X,s} |^2  +  | {\cal M}_{X,t}|  ^ 2 +  | {\cal M}_{X,u} |^2 \, . 
\label{M2D}
\eeq

\textit{2. The subtracted rate} ($\gamma_{\rm sub}$):
The subtracted rate is the difference between the full amplitudes~\eqref{eq:full_amp} and the amplitudes already contained in the D-graph~\eqref{M2D}, i.e.
\beq
|{\cal M}_{X,\rm sub}|^2 =  | {\cal M}_{X,\rm full} |^2 - | { \cal M}_{X,D} |^2 \, .
\label{M2sub}
\eeq

\textit{3. The top Yukawa rate} ($\gamma_{\rm top}$): 
This contribution differs from the previously discussed $2\to 2$ processes, as it focuses on gravitino production through scatterings involving top quarks.

The reason for performing this decomposition instead of computing directly the full squared amplitudes~\eqref{eq:full_amp} is that if a massless gauge boson is exchanged in the $t$- or $u$-channel, the corresponding squared matrix element exhibit infrared (IR)
divergences. This occurs in processes B,  F,  G and H of table~\ref{table:1}. The approach described above regularizes these IR divergences. This method was first used in~\cite{Rychkov:2007uq} for gravitinos, and in~\cite{Strumia:2010aa} and~\cite{Salvio:2013iaa} for axions (see also~\cite{Bouzoud:2024bom} for an alternative approach) and axinos, respectively.
We employed the same procedure in our previous study~\cite{Eberl:2020fml}, 
though the squared amplitudes for some processes differ by certain factors.
However, the subtracted components that impact our numerical results remain the same.

%%%%%%%%%%
%%%%%%%%%%

\section{The \texorpdfstring{2$\rightarrow$2}{2-->2} processes and their subtracted part}
\label{2to2sub}
In this section we will discuss the analytical results for the 
2$\rightarrow$2 processes involved in the gravitino production. Following the 
discussion in the previous section we also present the corresponding results 
for the subtracted part. 

The full amplitude squared is given as 
\beq
 |{\cal M}_{\rm full}|^2 = \sum_{X= {\rm A}}^{\rm J} |{\cal M}_{X, {\rm full}}|^2 \, .
\eeq
The same holds for $|{\cal M}_{\rm sub}|^2$ and $|{\cal M}_{\rm D}|^2$.

As can be seen in table~\ref{table:1} for the case of the $\suthree$ gauge group the incoming particles $a$, $b$ and the outgoing  $c$ could be gluons $g$, gluinos $\sg$, quarks $q$ or/and squarks $\sq$. 
Similar  processes occur  in the context of the gauge groups  $SU(2)_L$ and $U(1)_Y$. The factor $Y_N$ used in table~\ref{table:1} is defined as
\beq
Y_N =    1 + \frac{ M_N^2}{  3 m^2_{3/2}} \, .
\label{eq:yfactor}
\eeq 
Here and thereafter, the index $N=1,2,3$  
will be the  gauge group index running over the gauge groups of the SM: $\uone$, $\sutwo$, and $\suthree$. 
 $M_N$  and $ m_{3/2}$ represent the gaugino and  the gravitino mass, 
respectively. 
In $Y_N$ the unity originates  from  the 3/2 gravitino 
components and the mass  term   from  the 1/2 goldstino part. 
The Casimir operators appear in this table\footnote{In the thermal bath we have to sum over all incoming and outgoing states, e.g. $\sum_{a, b, c} |f^{a b c}|^2 = 3 \times 8 = 24 \equiv C_3$.}
are $ C_N =\sum_{a,b,c} |f^{abc}|^2= N(N^2-1)=\lbrace 0,6,24\rbrace$ and
$C_N' =\sum^{\phi}_{ a,i,j} |T^a_{ij}|^2 = \lbrace   11,21,48 \rbrace$.
With 
$\sum^{\phi}_{ a,i,j}$ we denote the sum over all involved chiral multiplets and  group indices.
The $f^{abc}$ are group structure constants  and $T^a$ the  generators.
%, e.g. for $SU(3)_c$ we have 6 quarks and 6 antiquarks, left- and right-handed. 
Please note that the processes A, B and F of table~\ref{table:1} does not appear  in  $U(1)_Y$, since  $C_1=0$.
Moreover we assume that the   particles  $a$, $b$ and $c$ are massless. 

The calculation of each of these  amplitudes in table~\ref{table:1} consists of two parts:  the pure gravitino spin  3/2 part and the gaugino  1/2 part. 
  Following~\cite{Rychkov:2007uq} for the gravitino 
we use  the gravitino 
polarization sum
\beq
\Pi^{3/2}_{\m \n}(P) =  \sum_{i = \pm 3/2} \grav^{(i)}_\m \, \overline{\grav}^{(i)}_\n = -\frac{1}{2} \g_\m \slash{P} \g_\n -  \slash{P} g_{\m\n}  \, ,
\label{eq:polsum}
\eeq
where $\grav_\m$ is the gravitino spinor and $P$ its momentum. 
For the or the goldstino 1/2 component, 
we use  the 
non-derivative approach described in~\cite{Moroi:1995fs,Pradler:2006tpx}. 
We have proved that 
the result for the full squared amplitude is  the same, either in the derivative or the non-derivative approach~\cite{Casalbuoni:1988kv,Casalbuoni:1988qd,Lee:1998aw}.
\begin{table}[t!]
\begin{center}
$\begin{array}{cccc}
\hline \hline
     \rowcolor{gray!15} 
\,\,\,\,\,\,X\,\,\,\,\,\, & \,\,\,\,\,\,{\rm process}\,\,\,\,\,\, & \,\,\,\,\,\,|{\cal M}_{X,\rm full}|^2\,\,\,\,\,\, &   \,\,\,\,\,\,| {\cal M}_{X,\rm sub}|^2\,\,\,\,\,\,\\
\hline\\[-0.42cm]
{\rm A} & g g \to \sg \Gr & 2 C_N ( s + 2 t + 2 t^2/s)   & - 2 s C_N \\
% \hline\\[-0.42cm]
{\rm B} & g \sg \to g \Gr & - 4 C_N (t + 2 s + 2 s^2/t)  &   2 t C_N  \\
% \hline\\[-0.42cm]
{\rm C} & \sq g \to q \Gr & 4 s C'_N  &  0 \\
% \hline\\[-0.42cm]
{\rm D} & g q \to \sq \Gr & - 4 t C'_N  &  0 \\
% \hline\\[-0.42cm]
{\rm E} & \sq q \to g \Gr & - 4 t C'_N  &  0 \\
% \hline\\[-0.42cm]
{\rm F} & \sg \sg \to \sg \Gr &  4 C_N (s^2 + t^2 + u^2)^2/(s t u)   &  0 \\
% \hline\\[-0.42cm]
{\rm G} & q \sg \to q \Gr & - 8 C'_N (s + s^2/t)  &  0 \\
% \hline\\[-0.42cm]
{\rm H} & \sq \sg \to \sq \Gr & -4 C'_N (t + 2 s + 2 s^2/t)     &  0 \\
% \hline\\[-0.42cm]
{\rm I} & q \bar q \to \sg \Gr & - 4 C'_N (t + t^2/s)  &  0 \\
% \hline\\[-0.42cm]
{\rm J} & \sq \bar\sq \to \sg \Gr & 2 C'_N ( s + 2 t + 2 t^2/s) &  0 \\
\hline
\end{array}$
\caption{Squared matrix elements for gravitino 
production in terms of $ g_N^2  Y_N/\mplanck ^2$ assuming massless particles, 
$Y_N = 1 +M^2_N/(3 m^2_{3/2})$,  $C_N = \{0,6,24\}$ and $C'_N = \{11,21,48\}$ and 
 $N = 1,2,3$ runs over the SM gauge groups. 
$| {\cal M}_{X,\rm sub}|^2$ is given in the non-derivative approach, as discussed in the text. }
%% Here all factors as discussed before, are included.
% The ten  processes are given for SU(3) but the results are shown for all three SM groups.}
%
\label{table:1}
\end{center}
\end{table}

It is worth noting that the amplitudes for processes B,  F,  G and H develop infrared divergences. As discussed earlier, to remove these divergences, we separate the total scattering rate into two parts: the subtracted  and the $D-$graph part. 
Additionally, for processes with incoming or/and outgoing gauge bosons, we have explicitly checked the gauge invariance  of the full amplitudes  squared,   $|{\cal M}_{X,\rm full}|^2 $.
On the other hand, $|{\cal M}_{X,\rm sub}|^2$ is gauge dependent.  As pointed out in~\cite{Rychkov:2007uq}, splitting the amplitudes into resummed and non-resummed contributions violates gauge invariance. Therefore, a gauge dependence of the result is expected. 

\subsection{The squared amplitudes}
\label{subsec:sqamp}

Many of the interaction terms concerning the gravitino in the full SUGRA Lagrangian \cite{Wess:1992cp} are irrelevant for our analysis, since the considered centre of mass energy $\sqrt{s}$ is much lower than the Planck scale $\mplanck$, thus  some operators are suppressed at least by a factor $\sim 1/\mplanck$. The relevant interaction Lagrangian\footnote{In this context, $\phi$, $\grav$, $\chi$, $\lambda$, and $F$ represent the bosons, gravitino, fermions, gauginos, and field strength, respectively.} is
\begin{align}
\label{eq:grav_Lagr}
  \mathcal{L}^{(N) }_{3/2,\,\rm int} &= -\frac{i}{\sqrt{2}\mplanck}\overline{\grav}_\m S^\m_{\rm MSSM} +\text{h.c.} \nonumber
  \\ 
  &= -\frac{i}{\sqrt{2}\mplanck} \left[
  \mathcal D^{(N) }_{\m} \phi^{*i} \overline{\grav}_\n \g^\m \g^\n \chi_L^i
  -\mathcal D^{(N) }_{\m} \phi^{i} \overline{\chi}_L^i \g^\n \g^\m  \grav_\n
\right] \nonumber \\
& \hspace{0.45cm}-  \frac{i}{8\mplanck}\overline{\grav}_\m [\g^\rho,\g^\sigma] \g^\m
  \lambda^{(N)\, a} F_{\rho\sigma}^{(N)\, a}\, ,
\end{align}
where in the first line $S^\m_{\rm MSSM}$ denotes the contribution from MSSM to the supercurrent. In the following, we will use this to calculate the relevant squared amplitudes.

It is worth noting that by applying crossing symmetry, one can observe 
that out of the ten squared amplitudes, only five are independent; namely, A, C, F, G, and H. 
The remaining amplitudes, B, D, E, I, and J, can be derived from these using crossing symmetry. 
The explicit analytic forms of the amplitudes A, C, F, G, and H, 
along with their corresponding Feynman diagrams, are presented in Appendix~\ref{appendix:A}. 
Additionally, the involved vertices are described, and instructions are provided 
for deriving the squared amplitudes for the other processes, B, D, E, I, and J, using crossing symmetry.

We consider the (massless) vector particles in the  $R_\xi$  gauge with  $\xi = 1$.
To ensure accuracy, we have verified the results involving external vector particles 
by also employing the axial gauge. The subtracted part is gauge-dependent and is 
presented in the  $\xi = 1$  gauge. Additionally, the spin-1/2 component of the 
gravitino is calculated using both the derivative and non-derivative approaches,
with full results found to be consistent across methods. The subtracted part is
provided exclusively in the non-derivative approach.

\subsection{The subtracted rate}
\label{sec:sub_rate}

In the fourth column of table~\ref{table:1}, we present the subtracted part for each process, expressed in terms of $ g_N^2  Y_N/\mplanck ^2$ and assuming massless particles. This contribution includes the sum of the interference terms among the four types of diagrams ($s$, $t$, $u$, $x$) and the squared $x$-diagram. It is non-zero only for the processes A and B.
In~\cite{Rychkov:2007uq}, the subtracted part for processes H and J is also non-zero. We assume the authors used the squark-squark-gluino-goldstino Feynman rule from~\cite{Bolz:2000xi}, which lacks a $\gamma _5$ factor. In contrast, we use the correct Feynman rule from~\cite{Pradler:2006tpx}, leading to a zero result for processes H and J.

To calculate the subtracted part  of thermal  rate   
 $a\,b \rightarrow c\, \widetilde{G}$, we use the expression: 
 \beq
\begin{aligned}
 \gamma _{\mathrm{sub}} \equiv   \mathcal{C } = \frac{1}{(2\pi)^8} \int &  \frac{{\rm d^3} \mathbf{p_a}}{2E_a}  \, \frac{{\rm d^3} \mathbf{p_b}}{2E_b} \, \frac{{\rm d^3} \mathbf{p_c}}{2E_c} \, \frac{{\rm d^3} \mathbf{p}}{2E} \, \,  |{\cal M}|^2\,  f_a \, f_b \, (1 \pm f_c) \d^4(P_a + P_b - P_c  - P)   \, ,
\label{collision_term1}
\end{aligned}
\eeq
where $f_i$ are the usual Bose-Einstein or Fermi-Dirac distributions given by:
\beq
\label{eq:fBF}
f_{B/F} = {1 \over e^{E \over T} \mp 1}\, .
\eeq
 In the relevant temperature range, all particles except the gravitino are in thermal equilibrium. As implied in the equation above, the statistical factor  $f_{\Gr}$ for the gravitino is negligible, allowing us to approximate $1 -  f_{\Gr} \sim 1$. Additionally, backward reactions can be neglected,  as they are  proportional to $f_{\Gr}$. 

 A remark with respect to our notation is in order: we use the symbol $\gamma$ for the 
 thermal production rate and $\mathcal{C}$ for the same quantity, as  calculated  for 
 the $2\to 2$ processes   in Appendix~\ref{AppendixC}. Both represent 
 the same quantity, which  serves as the source term in the Boltzmann equation~\eqref{eq:boltzmann}.

Please note that by approximating 
$1 \pm f_c \sim 1$, we can solve equation~\eqref{collision_term1} analytically. 
However, in our calculation, we retain the $1 \pm f_c$ factor in the integral, necessitating a numerical computation of the collision term ${\cal C}$. 
The detailed numerical method is provided in Appendix~\ref{AppendixC}.
\begin{figure}[t!]
\centering
\includegraphics[width=0.7\textwidth]{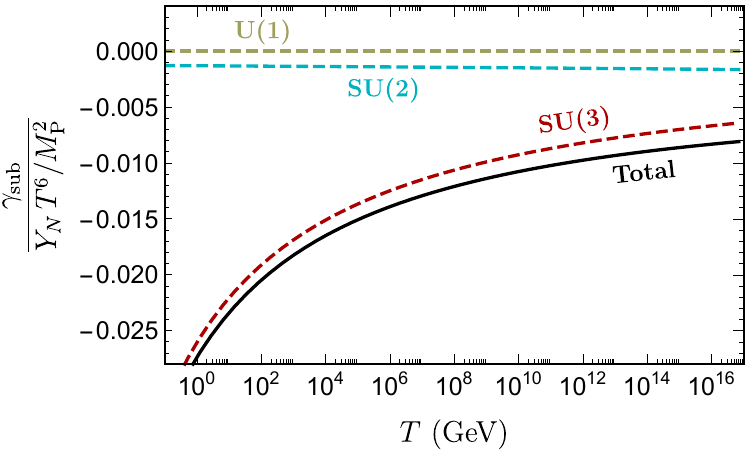}
\caption{The subtracted rate given by~\eqref{sub:part}  normalized by $Y_N T^6 / \mplanck ^2$. The curves represent, in order, the total subtracted rate and the contributions from $SU(3)_c$, $SU(2)_L$, and $U(1)_Y$.}
\label{Fig:sub_rate}
\end{figure}
If we use the amplitudes computed in the non-derivative approach (refer to Appendix~\ref{sec:non_der}), we see that the only nonzero contributions are those of processes A and B, which are
\begin{eqnarray}
&& |\mathcal{M}_{A, {\rm sub}}|^2 = \frac{1}{2} \frac{g_N^2}{\mplanck^2}\left(1+ \frac{M^2_N}{3 m^2_{3/2}} \right) (-2s C_N)\,, \label{sub_A}
\\ && |\mathcal{M}_{B, {\rm sub}}|^2 = \frac{g_N^2}{\mplanck^2}\left(1+  \frac{M^2_N}{3 m^2_{3/2}} \right) (2t C_N)\,.       \label{sub_B}
\end{eqnarray}
Note that in~\eqref{sub_A} a factor of $1/2$ is already included due to the 2 identical incoming  particles. Substituting~(\ref{sub_A}) and~\eqref{sub_B} in~(\ref{collision_term1}), the subtracted rate reads as  
\beq
\gamma _{\mathrm{sub}}=\frac{T^6}{\mplanck ^2} \sum_{N = 1}^3 g_N^2 \left(1+ \frac{M^2_N}{3 m^2_{3/2}} \right) C_N \left( -{\cal C}_{\mathrm{BBF}}^s +2{\cal C}_{\mathrm{BFB}}^t \right)\,.\label{sub:part}
\eeq
The numerical factors\footnote{Here we present only the numerical factors $ {\cal C}_{\mathrm{BBF}}^s $ and 
${\cal C}_{\mathrm{BFB}}^t$ required for this calculation. In Appendix~\ref{AppendixC} we also provide the coefficients for amplitudes proportional to $s^2$, $t^2$, and $st$ for various combinations of the involved particles. For more details, see table~\ref{tbl_camp}.}, calculated using the Cuba library~\cite{Hahn:2004fe}, are $ {\cal C}_{\mathrm{BBF}}^s = 0.260 \times 10^{-3} $ and 
${\cal C}_{\mathrm{BFB}}^t = -0.133 \times 10^{-3}\,. $ The subscripts B and F indicate whether the particles are bosons or fermions, respectively, while the superscripts denote whether the squared amplitude is proportional to $s$ or $t$. Our result for the subtracted part is negative, which is not unphysical because it is the total rate, not the subtracted part, that must be positive. In figure~\ref{Fig:sub_rate}, we show the total subtracted rate (black solid line) along with the contributions from individual gauge groups (colored dashed lines) as a function of temperature, spanning from $1$ to $10^{16}$ GeV.

We observe that the subtracted contribution $\g_{\mathrm{sub}}$ includes thermal corrections proportional to $T^6$.
Later, in the gravitino self-energy contribution ($D$-graph), we will implement thermal corrections to the gaugino and fermion propagators, which include higher orders of $T$ by performing a Dyson summation.

In the previous calculations, it was assumed that $\mgr$ is much smaller than $T$.  In the limit where $\mgr \to 0$, the gravitino interaction Lagrangian with the supercurrent $S_\m^{\rm MSSM}$ becomes
\beq
 \mathcal{L}_{3/2,\,\rm int} =\frac{- i }{\sqrt{2} \mpl }   \overline{\grav}^{\, \m} S_\m^{\rm MSSM} 
\xrightarrow[\mgr\to 0]{}   \frac{- i }{\sqrt{2} \mpl } \overline{\grav}^{\, \mu}_{(3/2)}  S_\m^{\rm MSSM} - \frac{i}{\sqrt{3}\mplanck} 
    \frac{\overline{\chi} (\partial^\m S_\m^{\rm MSSM})}{\mgr} \,,
    \label{eq:int}
\eeq
where it is assumed that in this limit, the gravitino decomposes as~\cite{Fayet:1977yc,Fayet:1977vd,Fayet:1979qi,Fayet:1979yb,Rychkov:2007uq}
\beq
\label{eq:grdecomp} 
\grav^\m  \xrightarrow[\mgr\to 0]{} \grav^\mu_{(3/2)}  -\frac{i}{\sqrt{6}} \gamma^\mu \chi -\sqrt{\frac{2}{3}} \frac{\partial^\mu \chi}{\mgr} \,, 
\eeq
and the goldstino coupling has been neglected.
The $\grav^\mu$ represents the full gravitino spinor and $\grav^\mu_{(3/2)}$ the  pure spin-3/2 gravitino components, while $\chi$ is the spin-1/2 goldstino component, both of which are massless in this limit.
The sum of the pure  gravitino components of $\grav^\mu_{(3/2)}$ is given in equation~\eqref{eq:polsum}.
This decomposition in equation~\eqref{eq:grdecomp} is the 
basis both for the derivative, and the non-derivative approaches presented in Appendices~\ref{sect:deriv} and ~\ref{sec:non_der}, respectively.

It should be noted that in~\cite{Rychkov:2007uq}, they approximate that the amplitude squared and the rate $\gamma$ decompose similarly, as
\beq
\g(\grav^\m) \simeq \g(\grav^\mu_{(3/2)}) + \g(\chi) \,. 
\eeq
The equality is not exact due to the omission of ``interference" terms, which are considered small.

\subsection{Top Yukawa contribution to the thermal rate}
\begin{figure}[t!] % H  
\centering{
\mbox{\includegraphics[trim={3.2cm 21.5cm 1cm 2.8cm},clip,scale=0.8]{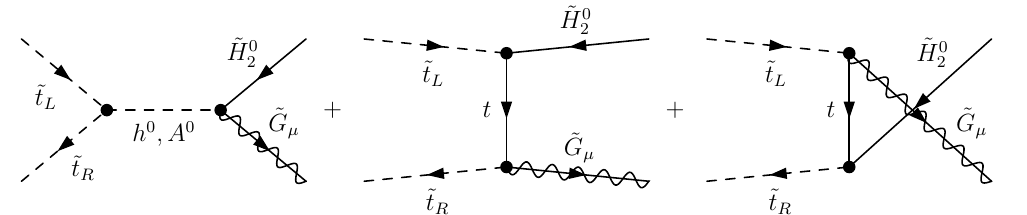}}  % trim={<left> <lower> <right> <upper>}, and clip must be activated
}
\caption[]{Feynman graphs for the process $\tilde t_L \tilde t^*_R \to \tilde H^0_2 \Gr$. 
\label{process_yt}}
\end{figure}

For the calculation of $\gamma_{top}$ we need the part of the amplitudes for\footnote{The particle $\tilde H_2^0$ denotes the superpartner of the Higgs field $H^0_2$, the  Higgsino.} $t \bar t \to \tilde H^0_2 \Gr$ 
and $\tilde t_i \tilde t^*_j \to \tilde H^0_2 \Gr$ which is proportional to the top Yukawa coupling $y_t$
in the limit of massless particles compared to the energy of the thermal bath. 
The contribution from $t \bar t \to \tilde H^0_2 \Gr$ vanishes and $\tilde t_i \tilde t^*_j$ becomes
$\tilde t_L \tilde t^*_R$ or $\tilde t_R \tilde t^*_L$. The Feynman graphs for the contributing amplitudes for $\tilde t_L \tilde t^*_R$
are shown in figure~\ref{process_yt}. 
Note that the graph with the four-point coupling of $\tilde t_R \tilde t^*_L  \tilde H^0_2 \Gr$ is 
zero in the $\xi = -2$ gauge for $\tilde G$.
The amplitudes for $\tilde t_L \tilde t^*_R \to \tilde H^0_2 \Gr$ are
\begin{eqnarray}
\label{A_H}
A^H & = & -  {\sqrt{2} \, c_F \over \mplanck} {y_t A_t \over s} \,  p_1^\mu  \bar u_{\mu}(p_2) P_{L} v(p_1) \, ,\\
\label{A_top}
A^{top} & = & -  {\sqrt{2} \,  c_F \over \mplanck} \,  y_t \left({k_1^\mu \over u} -  {k_2^\mu \over t} \right)  \bar u_{\mu}(p_2)  \slash{k}_1  P_{L}  v(p_1) \, ,
\end{eqnarray}
where $A^H$ corresponds to the left Feynman graph in figure~\ref{process_yt}, which is the sum of the graphs with $h^0$ and $A^0$  s-channel exchange,
and $A^{top}$ to the sum of the two other graphs in figure~\ref{process_yt}. The graph with $H^0$  s-channel exchange vanishes in that limit.
The colour factor $c_F$ is $\delta_{r s}$ with $r$ the colour of $\tilde t_L$ and $s$ the colour of $\tilde t^*_L$. Note, $|c_F|^2 = 3$, and $A_t$ is the trilinear soft breaking term in the stop sector.

We start from the amplitudes~\eqref{A_H} and~\eqref{A_top},
and calculate the squared amplitudes for $\tilde t_L \tilde t^*_R \to \tilde H_2^0 \Gr$.
For the spin~$\pm$3/2 contribution we use the gravitino polarization sum (\ref{eq:polsum}).
Only the two graphs with top propagator contribute,
\begin{equation}
 |A^{top}|^2  \to |A^{3 \over 2}|^2 = {3 y^2_t \over \mplanck^2}  s\, .
\end{equation}
For the spin~$\pm$1/2 contribution we work in the derivative approach, and apply the rule (\ref{equi_theo}) to~\eqref{A_H} and~\eqref{A_top}. Only the graphs with $h^0$ and $A^0$ propagators contribute,
\begin{equation}
|A^{H}|^2 \to |A^{1 \over 2}|^2  = {y^2_t \over \mplanck^2} {A^2_t \over m^2_{3/2}}  s\, .
\end{equation}
For  $\tilde t_R \tilde t^*_L \to \tilde H_2^0 \Gr$ we get the same results. Thus, the squared amplitude in order $y^2_t$ for 
the calculation of $\gamma_{top}$ is
\begin{equation}
|A|^2 = |A^{3 \over 2}|^2 + |A^{1 \over 2}|^2 = {6 y^2_t \over \mplanck^2} \left(1 + {A^2_t \over 3 m^2_{3/2}}  \right) s\, .
\end{equation}
This result is smaller by a factor of 12 compared to the one given in~\cite{Rychkov:2007uq}.
The top Yukawa rate, which is the collision term~\eqref{collision_term1}, with $|{\cal M}|^2 \equiv |A|^2$, is
\begin{equation}
\label{gamma:top}
\gamma_{top} =  {T^6 \over \mplanck^2} \, 6\, {\cal C}^s_{BBF}  y^2_t  \left(1 + {A^2_t \over 3 m^2_{3/2}}  \right) \, ,
\end{equation}
with ${\cal C}^s_{BBF}  = 0.260 \times 10^{-3}$, see table~\ref{tbl_camp}.

\subsection{Thermally corrected gravitino self-energy}

As outlined in Section 2, equation~\eqref{M2sub} describes the relation between the 
thermally corrected gravitino self-energy, the so-called  $D-$graph,
and the sum of squared amplitudes for the channels $s$, $t$, and $u$. 
In our analysis\footnote{In this section, we closely follow the notation of~\cite{Rychkov:2007uq}.} of the $D$-graph, we incorporate resummed thermal corrections to the gauge boson and gaugino propagators\footnote{Similar to~\cite{Rychkov:2007uq}, where the gravitino polarization sum~\eqref{eq:polsum} was utilized to nullify the corresponding quark-squark $D$-graph.}.
Although figure~\ref{D_GRAPH} specifically depicts the gluino-gluon thermal loop, our study encompasses contributions from all gauge groups.
The momentum flow used in calculating the $D-$graph is illustrated in figure~\ref{D_GRAPH}.
That is
    $\Gr(P) \to g(K) + \sg(Q)$,   with
\beq
 P = (p, p, 0, 0)\,, \,\,  K = (k_0, k \cos\theta_k, k \sin\theta_k, 0)\,,\,\, \text{and} \,\,\, Q = (q_0, q \cos\theta_q, q \sin\theta_q, 0) \,, 
 \label{eq:param_pkq}
\eeq
where $\theta_{k,q}$ are the polar angles of the corresponding 3-momenta $\mathbf{k}$ and $\mathbf{q} $ in spherical coordinates. Here,  we have already assumed that the gravitino is massless compared to the high temperature of the thermal bath, that is $P^2 = 0$.

We have seven variables:  $p, k, q, k_0, q_0, c_k,$ and $c_q$, and three non-trivial equations due to the overall momentum conservation, $P^i = K^i + Q^i,\, i = 0,1,2$. Thus, we are left with 
4 independent variables. Using the momentum conservation we obtain that
\begin{eqnarray}
p & = & k_0 + q_0\, ,  \\
\cos\theta _k & = & {p^2 - q^2 + k^2 \over 2 k p}\, , \label{c_k} \\
\cos\theta _q & = &  {p^2 + q^2 - k^2 \over 2 p q}\, ,  \label{c_q_prime} \\
\cos(\theta _k -\theta _q) & = &  {p^2 - q^2 - k^2 \over 2 k q}\, . \label{c_kmq_prime} \,
\end{eqnarray}

In order to calculate the gravitino selfenergy with vector-gaugino loop in the massless case 
we need the Feynman rules for the two vertices.
% The gluon-gluino-gravitino interaction is given by equation~\eqref{ } (see also~\cite{Eberl:2015dia}) and after obeying the equivalence theorem, the goldstino interaction can be read from~\eqref{  }.
Thus, the goldstino selfenergy with gluon-gluino loop including the outer goldstino legs can be written as
\begin{equation}
\Pi = {n_3 \over 8 \mplanck^2}  {M_{3}^2 \over 3 m^2_{3/2}} \int {{\rm d}^4 K \over (2 \pi)^4}
{\rm Tr}\left(\slash{P} [\slash{K}, \g^\m] S(Q)  [\slash{K}, \g^\n] D_{\m\n}(K)\right)\, , 
\end{equation}
in which $S(Q)$ is the gluino propagator,  $D_{\m\n}(K)$ the gluon propagator, and $n_3 = 8$ from the color running in the loop. 
Now, we can easily generalize that to the expression $\Pi^<$, including all three groups, $\uone$, $\sutwo$ and $\suthree$, with $n_1 = 1$ and $n_2 = 3$.
As we use the non-time ordered selfenergy $\Pi^<$, we get an additional factor of 1/2, that is 
% (see \eg \cite{Helmut-PhD} (B.7) middle line without the operator $T$)
\begin{equation}
\Pi^<(P) = {1 \over 16 \mplanck^2} \sum_{N=1}^3 n_N \left( 1 + {M^2_N \over {3 m^2_{3/2}}} \right) \int {{\rm d}^4 K \over (2 \pi)^4}
{\rm Tr}\left( \slash{P} [\slash{K}, \g^\m] \, {}^*S^<(Q)  [\slash{K}, \g^\n] \, {}^*D^<_{\m\n}(K) \right)\, , 
\label{def_Pi<}
\end{equation}
with the thermally resummed propagators denoted by a ${}^*$,
\begin{eqnarray}
{}^*S^<(Q) &=&{ f_F(q_0) \over 2} \left[ (\g_0 - \g^i \hat{q}^i) \rho_+(Q) +  (\g_0 + \g^i \hat{q}^i) \rho_-(Q)\right]\,,
\label{def_S<}\, \\ 
 {}^*D^<_{\m\n}(K) & = &  f_B(k_0) \left[ \Pi_{\m\n}^T \rho_T(K) + \Pi_{\m\n}^L  {k^2 \over K^2} \rho_L(K) + \xi {K_\m K_\n \over K^4}\right]\, ,
\end{eqnarray}
with $\hat{q}^i = q^i/q$. In~\eqref{def_Pi<} we have also incorporated the helicity $\pm 3/2$ components of the gravitino, as in~\cite{Bolz:2000fu} it has been shown, up to two loop order in the gauge couplings, that one obtains the characteristic factor $1+ M^2_N / {3  m^2_{3/2}}$, as well as in the calculation of the $2\to 2$ scatterings.
The spectral functions (longitudinal and transverse) for gauge-bosons are given by~\cite{Bellac:2011kqa,Kapusta:2006pm}
 \beq
 \rho_L=-2 \frac{P^2}{p^2}  {\rm Im}  \left(  \frac{1}{P^2-\pi_0-\pi_L}\right)\,, 
 \qquad
 \rho_T=-2{\rm Im} \left( \frac{1}{P^2-\pi_0-\pi_T} \right)\,,
 \label{eq:rhoLT}
 \eeq
 \begin{table}[t!]
\centering
\begin{tabular}{c c c c }
\hline \hline
     \rowcolor{gray!15} 
Gauge group &  $N_V$ & $N_F$ & $N_S$   \\
\hline  \\[-3.5mm]
\uone & $ 0 $ & $11$ & $22$ \\
\sutwo & $ 2 $ & $9$ & $14$ \\ 
\suthree & $ 3 $ & $9$ & $12$ \\
\hline
\end{tabular}
\caption{The numerical factors involved in the expressions for the vector-boson propagators. 
 Each value corresponds to a particular gauge group, \uone , \sutwo or \suthree, assuming the MSSM particle content. }
\label{table:Ns}
\end{table}
where the longitudinal-transverse thermal, and the $T=0$ corrections to the vector propagator are given by 
\begin{subequations}
\begin{align}
\pi_L =& - \frac{P^2}{p^2}g_N^2 (N_S H_S +N_F H_F + N_V H_V) \htlsim  - \frac{P^2}{p^2} (L+1) m_V^2\,, 
\\
\pi_T =& -\frac{\pi_L}{2} +\frac{g_N^2}{2} (N_S g_3 +N_F G_F + N_V G_V) \htlsim  \left(1+\frac{P^2}{p^2} \frac{L+1}{2}\right) m_V^2\,,
\\
\pi_0 =& g_N^2P^2 \frac{2N_F +N_S/2 -5N_V}{48\pi^2} \ln\left(- \frac{P^2}{\m^2}\right)\,.
\label{eq:pi_LT0}
\end{align}
\end{subequations}

In equations above,  the HTL refers to the hard thermal loop approximation, which is valid at energies and momenta $p, p_0 \ll T$. It accurately describes the thermal effects that occur at $p, p_0 \sim g_N T$ when $g_N \ll 1$. However, this approximation does not hold for physical gauge couplings, particularly for the strong coupling $g_3$.  Note also that $\pi_L$ and $\pi_T$ do not contain the one-loop quantum correction at $T=0$, so it must be added separately.

The integrals $H_{S,F,V}, G_{S,F,V}$ and the function $L$ can be found in Appendix~\ref{AppendixB}, while the numerical factors $N_{S,F,V}$ are given in table~\ref{table:Ns} assuming the MSSM particle content. The vector thermal mass is given by 
\begin{equation}
m_V^2= \frac{1}{6}g_N^2 T^2 (N_V+N_S/2+N_F/2)\,.
\label{thermalmass}
\eeq
For fermion particles ($ \r_{+} $) and fermion holes ($ \r_{-} $) the spectral functions, including the $T=0$ contribution, are given by
\beq
\r_\pm  =  -{\rm Im} \left\{  \Big(   \o_\mp\left[ 1 + \frac{1}{ 2 \pi^2} \frac{m_F^2}{T^2} \left(2  - \ln{ (- \frac{P^2}{ \mu^2} ) } \right) \right]
+  {m_F^2 \over T^{2}} F_\pm  \Big)^{-1}  \right\} \,, \quad \o_\pm = \frac{p_0\pm p}{2}\,, 
\label{eq:rhopm}
\eeq
where the thermal fermion mass is 
\begin{equation}
m_F^2 = \Big({g_N^2 C_R \over 8}  + {\l^2_q \over 16}\Big) T^2\, ,
\end{equation}
and
\begin{align}
F_\pm  =  & \mp \int_0^\infty {{\rm d} k \over \pi^2} \bigg[  {\o_\mp \over p^2}\Big( k L_+ (n_B + n_F) + L_- \left( n_B \o_\mp + n_F  \o_\pm\right)\Big)  \nn
& + 2   {L \o_\mp + \o_\pm \over p p_0} k (n_B + n_F)\bigg] \,  
\quad  \htlsim  \quad   \mp \frac{T^2}{2}{L \o_\mp + \o_\pm \over p p_0}.
\label{eq:Fpm}
\end{align}
The statistical factors $n_{B,F}$ are given by~\eqref{eq:fBF} by substituting $E \to |E|$, while the group factor $C_R = \{1, 3/4, 4/3 \}$~\cite{Weldon:1982bn} and $\lambda_q$ is the scalar-fermion-fermion coupling (see equation~\eqref{eq:scferfer}).

Both the vector and fermion spectral functions\footnote{Note that the temperature dependence of the spectral functions is 
$\rho_{L,T}   \sim T^{-2}$ and  $\rho_{\pm}  \sim T^{-1}$. } can be decomposed as
\beq
\rho_{L,T} = \rho_{L,T}^{\rm pole} + \rho_{L,T}^{\rm cont} , \quad \rho_{\pm} = \rho_{\pm}^{\rm pole} + \rho_{\pm}^{\rm cont} ,
\eeq
where the continuum contribution is computed directly from the definitions in equations~\eqref{eq:rhoLT} and~\eqref{eq:rhopm}. To compute the pole contribution, one must solve the dispersion relations to determine the pole positions and calculate the residues at these poles. In~\cite{Rychkov:2007uq}, the HTL approximation was assumed to obtain analytical results for the pole positions and residues. In our work, however, we utilize numerical methods to solve the full one-loop dispersion relations (as detailed in Appendix~\ref{sec:disp_rel}), without relying on the HTL approximation. This approach allows us to compute both the pole positions and residues using the complete one-loop thermal corrections.

Using the parametrization~\eqref{eq:param_pkq} and the abbreviations ``c" and ``s" for cosine and sine, we get for the different spectral function combinations of equation~\eqref{def_Pi<} the following results:
\begin{align}
\label{eq:-L}
\propto  \rho_- \rho_L:   &  16\, p\, k^2 ( c_{2k} c_q + s_{2k} s_q + 1)\, ,\\
\propto  \rho_+ \rho_L:  & 16\, p\, k^2 (-c_{2k} c_q - s_{2k} s_q + 1)\, ,\\
\propto  \rho_- \rho_T:   &  16\, p \left((k_0^2 + k^2)(2-\cos(2\theta _k -\theta _q)- c_q  ) - 4 k_0 k (c_k-\cos(\theta _k -\theta _q))\right)\, ,\\
\propto  \rho_+ \rho_T:  &  16\, p \left((k_0^2 + k^2)(2+\cos(2\theta _k -\theta _q)+ c_q  ) - 4 k_0 k (c_k + \cos(\theta _k -\theta _q))\right)\, .
\label{eq:+T}
\end{align}
% Note, that $c_{2k} c_q + s_{2k} s_q + 1 = 2 \cos^2{2\theta_k - \theta_q \over 2}$ and
% $-c_{2k} c_q - s_{2k} s_q + 1 = 2 \sin^2{2\theta_k - \theta_q \over 2}$. 
% Substituting these in equation~\eqref{def_Pi<} we get
After some algebra equation~\eqref{def_Pi<} becomes
\begin{align}
\label{Pi<1dp}
\Pi^<(&P)  = {1 \over 32 \mplanck^2} \sum_{N=1}^3 n_N \left(1 +{M^2_N \over 3 m^2_{3/2}}\right) \int {{\rm d}^4 K \over (2 \pi)^4}  f_F(q_0)  f_B(k_0) \times \\ 
& \bigg[ \rho_L (K) \rho_- (Q) 32\, p\, k^2 \cos^2{2\theta_k - \theta_q \over 2} + \rho_L (K) \rho_+ (Q)   32\, p\, k^2 \sin^2{2\theta_k - \theta_q \over 2} \nn 
& + \rho_T (K) \rho_- (Q) 16\, p \left((k_0^2 + k^2)(2-cos(2\theta _k -\theta _q)- c_q  ) - 4 k_0 k (c_k-cos(\theta _k -\theta _q))\right) \nn
& +  \rho_T (K) \rho_+ (Q) 16\, p \left((k_0^2 + k^2)(2+cos(2\theta _k -\theta _q)+ c_q  ) - 4 k_0 k (c_k + cos(\theta _k -\theta _q))\right)\bigg]\, . \nonumber
\end{align}
 In order to compute the integral~\eqref{Pi<1dp} it is convenient to multiply by the $4$-momentum $ \delta $-function $ \int {\rm d}^4Q\, \delta^4(K+Q-P)=1 $. Also, using the relations
\begin{equation}
\begin{aligned}
 {\rm d}^4K &= {\rm d}k_0\,k^2 {\rm d}k\, {\rm d}\cos\theta_k\,{\rm d}\phi_k \quad \text{and} \quad  {\rm d}^4Q &= {\rm d}q_0\,q^2 {\rm d}q\, {\rm d}\cos\theta_q\,{\rm d}\phi_q\,,
\label{d4KQ}
\end{aligned} 
\end{equation}
we can perform the integrations $ {\rm d}q_0 $, $ {\rm d}\cos\theta_q $, ${\rm d}\cos\theta_k $ and $ {\rm d}\phi_k $ thanks to the $ \delta $-function.
\begin{figure}[t]
\centering
\includegraphics[width=0.6\textwidth]{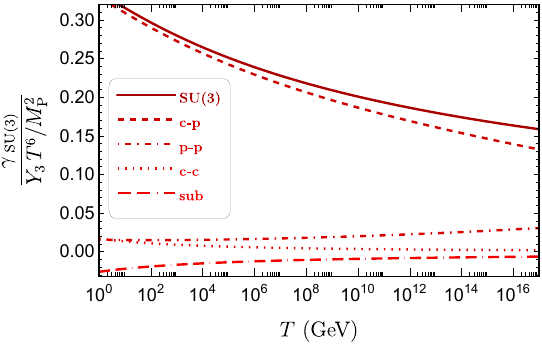}
\caption{Various contributions to the $SU(3)_c$ production rate, normalized by $Y_N  T^6 / \mplanck^2$, are shown. The solid line represents the total result, while the other lines correspond to the ``cont-pole", ``pole-pole", ``cont-cont"  and subtracted contributions, respectively.} 
\label{fig:gamma_etc}
\end{figure}
Nothing depends on $ {\rm d}\phi_k $, so we get an extra $2\pi$ from this integration. After these integrations the equations~\eqref{eq:-L}-\eqref{eq:+T} become,
\begin{align}
\propto  \rho_- \rho_L:   & \frac{8}{q} (p - q)^2  ((p + q)^2 - k^2)\,, \\ 
\propto  \rho_+ \rho_L:  & \frac{8}{q} (p + q)^2 (k^2 - (p - q)^2) \, ,\\  
\propto  \rho_- \rho_T:   &  \frac{8}{q} \left(k^2 - (p - q)^2\right)\left( \left(1 + {k_0^2 \over k^2} \right) \left( k^2 + (p + q)^2\right) 
- 4 k_0 (p + q)\right) \, ,\\
\propto  \rho_+ \rho_T:  &  \frac{8}{q} \left( (p + q)^2 - k^2\right)\left( \left(1 + {k_0^2 \over k^2} \right) \left( k^2 + (p - q)^2\right)  
- 4 k_0 (p - q)\right)\, ,
\label{ptdp}
\end{align}
and the resummed propagator~\eqref{Pi<1dp} takes the form
\begin{align}
\hspace*{-1mm}
\Pi^<(&P) = {1 \over 4(2\pi)^3}{1 \over \mplanck^2} \sum_{N=1}^3 n_N \left(1\! + \!{M^2_N \over 3 m^2_{3/2}}\right) \int_{-\infty}^\infty \!{\rm d}k_0  \int_0^\infty \!{\rm d}k \int_{|k-p|}^{k+p} \!{\rm d}q\, \frac{k}{p} f_B(k_0)f_F(p_0\!-\!k_0) \nn
& \hspace{-0.38cm}\times   \bigg[ \rho_L (K) \rho_- (Q) (p - q)^2  ((p + q)^2 - k^2) + \rho_L (K) \rho_+ (Q) (p + q)^2 (k^2 - (p - q)^2)\nn
& \hspace{-0.38cm}+ \rho_T (K) \rho_- (Q) \left(k^2 - (p - q)^2\right)\left( \left(1 + {k_0^2 \over k^2} \right) \left( k^2 + (p + q)^2\right) - 4 k_0 (p + q)\right)  \nn
& \hspace{-0.38cm}+  \rho_T (K) \rho_+ (Q) \left( (p + q)^2 - k^2\right)\left( \left(1 + {k_0^2 \over k^2} \right) \left( k^2 + (p - q)^2\right) - 4 k_0 (p - q)\right)\bigg]\, , 
\label{Pi<1dp2}
\end{align}
with $ q_0=p-k_0$. In order to compute the production rate $ \gamma_D $ we will use its definition 
\cite{Bellac:2011kqa, Kapusta:2006pm}
\begin{equation}
\gamma_D = \int{{\rm d^3 \mathbf{p}} \over 2 p_0 (2 \pi)^3} \Pi^<(p)\,, 
\label{def_gamma_D_dp}
\end{equation}
with $ {\rm d}^3 \mathbf{p}=4\pi p^2{\rm d}p\, $ in this frame. Then
\begin{align}
\gamma_D & = {1 \over 4(2\pi)^5}{1 \over \mplanck^2} \sum_{N=1}^3 n_N \left(1 + {M^2_N \over 3 m^2_{3/2}}\right) \times
\nonumber\\
& \hspace*{3cm} \int_0^\infty {\rm d}p \int_{-\infty}^\infty {\rm d}k_0  \int_0^\infty {\rm d}k \int_{|k-p|}^{k+p} {\rm d}q\, k f_B(k_0)f_F(p_0-k_0)\times \nn
& \hspace{-0.2cm}\bigg[ \rho_L (K) \rho_- (Q) (p - q)^2  ((p + q)^2 - k^2) + \rho_L (K) \rho_+ (Q) (p + q)^2 (k^2 - (p - q)^2)\nn
& \hspace{-0.2cm}+ \rho_T (K) \rho_- (Q) \left(k^2 - (p - q)^2\right)\left( \left(1 + {k_0^2 \over k^2} \right) \left( k^2 + (p + q)^2\right) - 4 k_0 (p + q)\right)  \nn
& \hspace{-0.2cm}+  \rho_T (K) \rho_+ (Q) \left( (p + q)^2 - k^2\right)\left( \left(1 + {k_0^2 \over k^2} \right) \left( k^2 + (p - q)^2\right) - 4 k_0 (p - q)\right)\bigg]\, .
\label{gammadgraph}
\end{align}
\begin{figure}[t]
\centering
\includegraphics[width=0.6\textwidth]{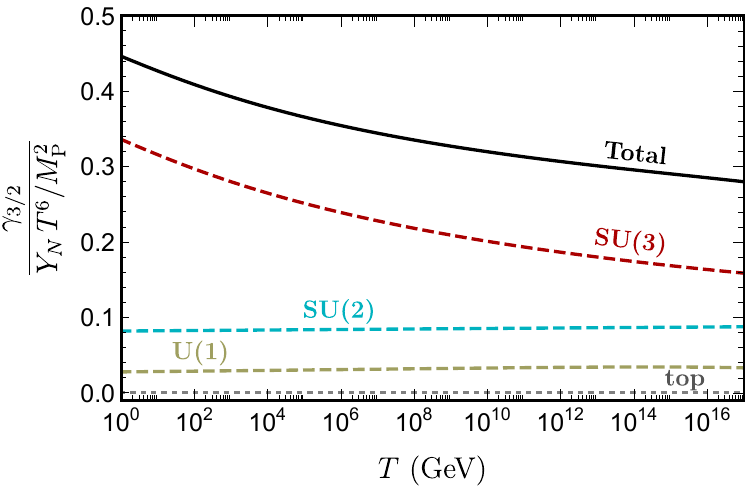}
\caption{The gravitino production rate normalized by $Y_N T^6 / \mplanck ^2$. The curves represent, in order, the total rate given by~\eqref{eq:gamma32}, the $SU(3)_c$, $SU(2)_L$, and $U(1)_Y$ contributions given by~\eqref{gamma:paramtrization}, and the top Yukawa rate given by~\eqref{gamma:top}. The top Yukawa coupling  is  $y_t=0.7$.} 
\label{fig:gamma32}
\end{figure}
The spectral functions $\rho_{L,T}$ and $\rho_{\pm}$ are provided in equations~\eqref{eq:rhoLT} and~\eqref{eq:rhopm}. The thermally corrected one-loop self-energies for gauge bosons, scalars, and fermions, utilized in the calculation of these spectral functions, are detailed in Appendices~\ref{Vector-boson self-energy} and~\ref{Fermion self-energy}, and are also available in various previous works~\cite{Weldon:1982aq, Weldon:1982bn, Weldon:1989ys, Weldon:1996kb, Peshier:1998dy, Weldon:1999th}. Comparing our result with that of~\cite{Rychkov:2007uq}, we observe differences in the overall factor and the number of independent phase-space integrations. Our analytical result has been verified using various frames for the momentum flow into the loop. The four dimensional integral~\eqref{gammadgraph} has been calculated numerically using the Cuba library~\cite{Hahn:2004fe}.
In this computation, the products of the different spectral densities yield contributions of the forms ``pole-pole", ``cont-pole", and ``cont-cont". Consequently, we must evaluate integrals of dimensions two, three, and four, respectively, due to the presence of $\delta$-functions associated with the pole parts of the spectral densities. Figure~\ref{fig:gamma_etc} illustrates the various contributions to the production rate for the $\suthree$ group. Among these, the ``cont-pole" contribution is the most significant, surpassing the other contributions. In figure~\ref{fig:gamma32} we show the full gravitino production rate (black solid line) given by equation~\eqref{eq:gamma32} in terms of the factor $Y_N T^6 / \mplanck ^2$. The dashed lines represent the individual contributions from each gauge group, while the gray dotted line shows the top quark contribution\footnote{Note that the top quark contribution in figure~\ref{fig:gamma32} is normalized by $(1 + {A^2_t / 3 m^2_{3/2}})T^6 / \mplanck ^2$.}
given by equation~\eqref{gamma:top}.
Compared to previous works that implement the same method for computing the gravitino production rate~\cite{Rychkov:2007uq,Eberl:2020fml}, our updated results show a decrease of approximately $25-30\%$ across the entire temperature range when compared to~\cite{Rychkov:2007uq}. Relative to~\cite{Eberl:2020fml}, we observe a reduction of about $40\%$ at $T \sim 1\GeV$. This reduction decreases with rising temperatures, reaching around $20\%$ at $T \sim 10^8\GeV$, and disappears entirely at the grand unification (GUT) scale $2\times10^{16}\GeV$.

In the following section, we will introduce a useful parametrization of this integral, inspired by~\cite{Ellis:2015jpg}. This parametrization incorporates the subtracted rate calculated in section~\ref{sec:sub_rate}, providing a more refined approach to the integral's evaluation.

\section{The total gravitino thermal production rate}
\label{sec:tot_res}
Since the computation of the gravitino production rate presented in figure~\ref{fig:gamma32} is purely numerical, it would be useful to have some practical parametrizations. Ellis et al.~\cite{Ellis:2015jpg} were the first to parametrize the results of~\cite{Rychkov:2007uq} using the gauge coupling constants $g_N$, and we applied a similar approach in~\cite{Eberl:2020fml}. Therefore, we have parametrized the results for the subtracted part and the D-graph using the gauge couplings $g_Y$, $g_2$, and $g_3$. Specifically,
\beq 
 \gamma_{\mathrm{sub} }+ \gamma_{\mathrm{D}} =   {3  \zeta(3)\,  T^6 \, \over 16  \pi^3 \mplanck ^2}   \sum_{N = 1}^3  c_N \, g_N^2 \,  \left(1 + {M^2_N \over 3 m^2_{3/2}}\right) \ln\left( \frac{k_N}{g_N}\right)\,,
\label{gamma:paramtrization}
\eeq
where the constants $c_N$ depend on the gauge group and their values are given in table~\ref{table:param} (left).
By adding the contribution proportional to  $h_t$,  we obtain the total gravitino thermal production rate
\beq 
\gamma_{3/2} = \frac{T^6}{\mplanck ^2} \left[ {3  \zeta(3)\,  \over 16  \pi^3 }   \sum_{N = 1}^3  c_N \, g_N^2 \,  \left(1 + {M^2_N \over 3 m^2_{3/2}}\right) \ln\left( \frac{k_N}{g_N}\right) +6\, {\cal C}^s_{BBF}  y^2_t  \left(1 + {A^2_t \over 3 m^2_{3/2}}  \right) \right]\,,
\label{eq:gammatotpar}
\eeq
where from  table~\ref{tbl_camp} we get ${\cal C}^s_{BBF}  = 0.260 \times 10^{-3}$.
The  detailed calculation of this factor can be found in Appendix~\ref{AppendixC}.
It is worth noting that 
$\gamma_{3/2}$ is related to the thermalized cross section 
$\left\langle \sigma  \, v_{\rm rel} \right\rangle$, as
\beq
\gamma_{3/2}= \left\langle \sigma\,  v_{\rm rel} \right\rangle n_{\rm rad}^2\, ,
\label{eq:sigmav}
\eeq
where $n_{\rm rad}= \zeta(3)T^3/\pi^2$.

The above parametrization can be rewritten in a more convenient form as a function of temperature rather than coupling constants, as follows:
%%%%
\begin{table}[t!]
\centering
\begin{tabular}{c c c }
\hline \hline
     \rowcolor{gray!15} 
Gauge group &  $c_N$ & $k_N$   \\
\hline  \\[-3.5mm]
\uone & $ 35.56 $ & $0.85$ \\
\sutwo & $ 35.33 $  & $1.38$\\ 
\suthree & $ 29.40 $  & $3.07$ \\
\hline
\end{tabular}
\qquad\qquad
\begin{tabular}{c c c c }
\hline \hline
     \rowcolor{gray!15} 
Gauge group &  $\xi_N$ & $\sigma_N$ & $\tau_N$   \\
\hline  \\[-3.5mm]
\uone & $ 0.93 $ & $60.38$ & $2.30$ \\
\sutwo & $ 10.14 $  & $188.33$ & $3.72$\\ 
\suthree & $ 2.81 $  & $12.73$ & $1.02$ \\
\hline
\end{tabular}
\caption{The values of the constants $c_N, k_N$ (left) and  $\xi_N, \sigma_N, \tau_N$ (right) that parametrize our result for the subtracted and the D-graph part. Each value corresponds to a particular gauge group, \uone , \sutwo, and \suthree . }
\label{table:param}
\end{table}
%%%%%
\beq
\gamma_{\mathrm{sub} }+ \gamma_{\mathrm{D}}  = \sum_{N = 1}^3 \frac{T^6}{\mplanck ^2}\left(\xi_N\frac{\ln\left(\sigma_N\pm \ln{T}\right) -\tau_N}{\sigma_N\pm \ln{T}} \right) \left(1 + {M^2_N \over 3 m^2_{3/2}}\right)\,,
\label{gamma:paramtrization_2}
\eeq
where the plus sign corresponds to $\suthree$ and the minus sign to $\sutwo$ and $\uone$ respectively. The temperature $T$ is measured in$\GeV$ and the constants $\xi_N, \sigma_N$ and $\tau_N$ are given in table~\ref{table:param} (right). 
The above equation for the total width $\gamma_{3/2}$ reads as
\beq 
\gamma_{3/2} = \frac{T^6}{\mplanck ^2} \left[\sum_{N = 1}^3 \left(\xi_N\frac{\ln\left(\sigma_N\pm \ln{T}\right) -\tau_N}{\sigma_N\pm \ln{T}} \right) \left(1 + {M^2_N \over 3 m^2_{3/2}}\right)  +6\, {\cal C}^s_{BBF}  y^2_t  \left(1 + {A^2_t \over 3 m^2_{3/2}}  \right) \right]\,.
\label{eq:gammatotpar_2}
\eeq

In~\eqref{gamma:paramtrization_2}  we did not introduce a new parametrization. Instead, we substituted the leading-order
logarithmic corrections to the running of all the gauge couplings~\cite{Martin:1997ns}, assuming a unified gauge coupling $g_{\rm GUT} = \sqrt{\pi/6}$ at the GUT scale. Note also that the gaugino masses are temperature-dependent and are assumed to unify at the same GUT scale, with the values $M_{N} = M_{1/2} \left(g_N^2/g_{\rm GUT}^2\right)$.

\section{Gravitino abundance, DM and reheating temperature}
\label{sec:abund}
The gravitinos are produced after inflation mainly by three mechanism: through decays of 
the inflaton field~\cite{Kallosh:1999jj,Giudice:1999am,Nilles:2001ry,Kawasaki:2006gs,Endo:2006qk,Ellis:2015jpg,Dudas:2017rpa,Kaneta:2019zgw},  directly non-thermally through decays of heavier  unstable supersymmetric particles, like gauginos or sfermions~\cite{Giudice:1998bp,Choi:1999xm,Asaka:2000zh,Jedamzik:2005ir,Fukushima:2013vxa} and last  but not least 
thermally through $1\to 2 $ or $2 \to 2$ processes with a gravitino in 
the final state~\cite{Weinberg:1982zq,Ellis:1984eq,Khlopov:1984pf,Moroi:1993mb,Kawasaki:1994af,Moroi:1995fs,Ellis:1995mr,Bolz:1998ek,Bolz:2000fu,Bolz:2000xi,Steffen:2006hw,Pradler:2006qh,Pradler:2006hh,Rychkov:2007uq,Pradler:2006tpx,Ellis:2015jpg, Eberl:2020fml}. Apparently, for the later the most important ingredient is the gravitino thermal production rate 
calculated in equation~\eqref{gamma:paramtrization}.

In order to obtain the thermal gravitino abundance one has to solve 
the Boltzmann equation for the gravitino number density $n_{3/2}$ 
\beq
\frac{{\rm d} {n}_{3/2}}{{\rm d}t} + 3H n_{3/2} = \gamma_{3/2}\,,
\label{eq:boltzmann}
\eeq
where $H$ is the Hubble constant. The gravitino abundance $Y_{3/2}$ is defined as
\beq
Y_{3/2}= {n_{3/2} \over n_{\rm rad} }\,,
\label{eq:abundance}
\eeq
where we have used for the radiation density  $n_{\rm rad}=\zeta(3)T^3/\pi^2$. Substituting $Y_{3/2}$
into~\eqref{eq:boltzmann} we get  the Boltzmann equation for $Y_{3/2}$
\beq
\frac{{\rm d} Y_{3/2}}{{\rm d}t}  + 3\left(H+ \frac{1}{T}\, \frac{{\rm d}T }{{\rm d} t}\right)Y_{3/2} =  \frac{\gamma_{3/2}}{ n_{\rm rad}}\, .
\label{eq:Y_boltz}
\eeq
%%%%
Using the  entropy conservation relation
\beq
\frac{{\rm d}( g_{*s} T^3a^3)}{{\rm d}t}=0 \, , 
\eeq 
where  $g_{*s}$ are the effective entropy degrees of freedom and $a$ the scale factor,  we can trade time with temperature in equation~\eqref{eq:Y_boltz}
as
\beq
\frac{{\rm d}Y_{3/2}}{{\rm d}T} - Y_{3/2}\,  \frac{{\rm d}\ln g_{*s} }{{\rm d}T} = 
       -\frac{  \gamma_{3/2}  } {H\, T\,  n_{\rm rad}} \left[1+\frac{T}{3} \frac{{\rm d}\ln g_{*s}  }{{\rm d}T}\right]\,.
\label{eq:Y_boltz2}
\eeq
By integrating  equation~\eqref{eq:Y_boltz2}  
from  $T_{\rm reh}$ down  to temperature $T$ and assuming 
  that inflaton decay and thermalization are instantaneous and simultaneous at $T_{\rm reh}$, 
one gets   
\bea
Y_{3/2}(T)&=&Y_{3/2}(T_{\rm reh})\,\frac{g_{*s}(T)}{g_{*s}(T_{\rm reh})} \nonumber \\
&& - g_{*s}(T)\int_{T_{\rm reh}}^T {\rm d}\tau \, \frac{\gamma_{3/2} (\tau)} {g_{*s}(\tau)\, H(\tau)\, n_{\rm rad}(\tau)\,\tau}\left[1+\frac{\tau}{3} \frac{{\rm d}\ln g_{*s}(\tau)}{{\rm d}\tau}\right]\,.
\label{eq:Y_integr}
\eea
Furthermore, applying the standard freeze-in scenario assuming negligible initial  gravitino 
abundance at $T_{\rm reh}$,    $Y_{3/2}(T_{\rm reh})\simeq 0$,   and ignoring 
a weak temperature dependence of the integrand in the r.h.s. of~\eqref{eq:Y_integr}, 
 we get that the gravitino abundance in the limit $T \ll T_{\rm reh}$ reads as~\cite{Ellis:2015jpg}
\beq
Y_{ 3/2}(T)   \underset{T \ll T_{\rm reh}}{\longrightarrow}    {\gamma _{3/2}(T_{\rm reh}) \over H(T_{\rm reh}) \,\,  n_{\rm rad}(T_{\rm reh}) }\,\, {g_{*s}(T) \over g_{*s}(T_{\rm reh})}\, .
\label{eq:abundance_app}
\eeq
We notice that in~\cite{Ellis:2015jpg,Garcia:2017tuj}, it has been  estimated that the 
approximation of the 
instantaneous inflaton decays and the thermalization can affect the result of the
 gravitino DM abundance by  $\sim 10\%$.

Moreover, using the definitions 
\beq
\Omega _{3/2} =\frac{\rho_{3/2}}{ \rho _{\rm cr}}\,   , \qquad  \rho_{3/2} = n_{3/2}  m_{3/2}\,,
\label{eq:definitions}
\eeq
 and the equations~\eqref{eq:abundance}, \eqref{eq:abundance_app} and  \eqref{eq:gammatotpar},  one  gets $\Omega _{3/2} h^2 $ as
 \bea
\Omega _{3/2} h^2 &=& {\rho_{3/2}(t_0) h^2 \over \rho _{\rm cr} } = {m_{3/2} Y_{3/2}(T_0)n_{\rm rad}(T_0) h^2 \over \rho _{\rm cr} }
  \simeq  1.33\times 10^{24} {m_{3/2} \gamma_{3/2}(T_{\rm reh}) \over T_{\rm reh}^5 } \nn
  &\simeq & 0.016 \left( \frac{m_{3/2}}{1\TeV}\right)\left(\frac{T_\mathrm{reh}}{10^{10} \GeV} \right) \Bigg[\sum_{N = 1}^3 \, c_N \, g_N^2 \,  \left(1 + {M^2_N \over 3 m^2_{3/2}}\right) \ln\left( \frac{k_N}{g_N}\right) \nn
  && + 0.214 y^2_t  \left(1 + {A^2_t \over 3 m^2_{3/2}}  \right) \Bigg]\,.
\label{eq:dmdensity}
\eea
Here, $\rho _{\rm cr}=3\, H_0^2 \mplanck ^2$ is the critical  energy density,  $H_0=100\, h\, {\rm km/(s\, Mpc)}$ is
the Hubble constant,  and $T_0 =2.7 K$ is the current temperature of the cosmic microwave background.   The entropy degrees of freedom at the relevant temperatures are $g_{*s}(T_0)=43/11$ and $g_{*s}(T_{\rm reh})=915/4$. The latter value corresponds to the effective energy degrees of freedom for  $H(T_{\rm reh})$   in the MSSM.

Following the latest data from the Planck satellite, the cosmological accepted value for the 
DM density in the Universe  is
 $\Omega _\mathrm{DM} h^2= 0.12 \pm 0.0012$~\cite{Aghanim:2018eyx}.
Assuming that the thermal gravitino abundance accounts for the total 
observed DM in the Universe, we can derive bounds on the 
gravitino mass and/or the reheating temperature. 
However, as we will see, these bounds significantly weaken when
allowing for other particles to play the role of the DM particle.

\begin{figure}[t!] % H
\centering
{\includegraphics[width=0.485\textwidth]{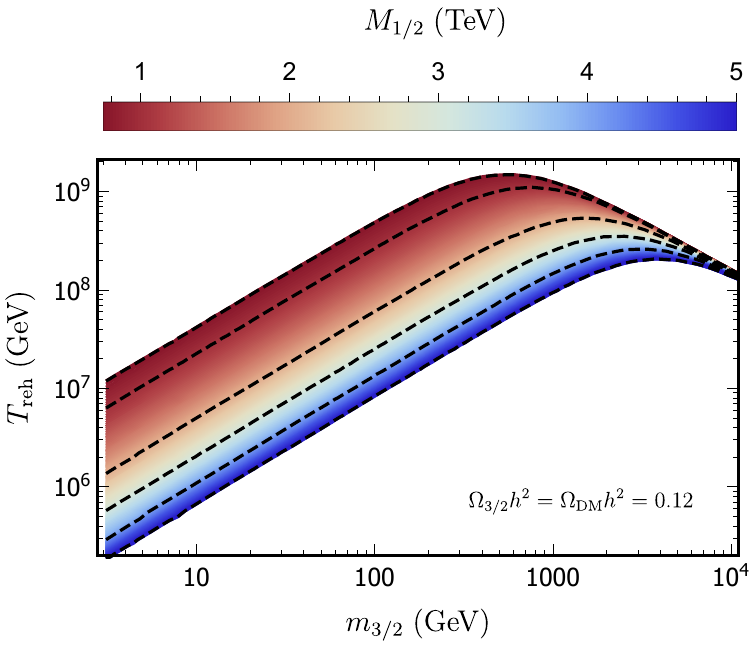} \,
\includegraphics[width=0.485\textwidth]{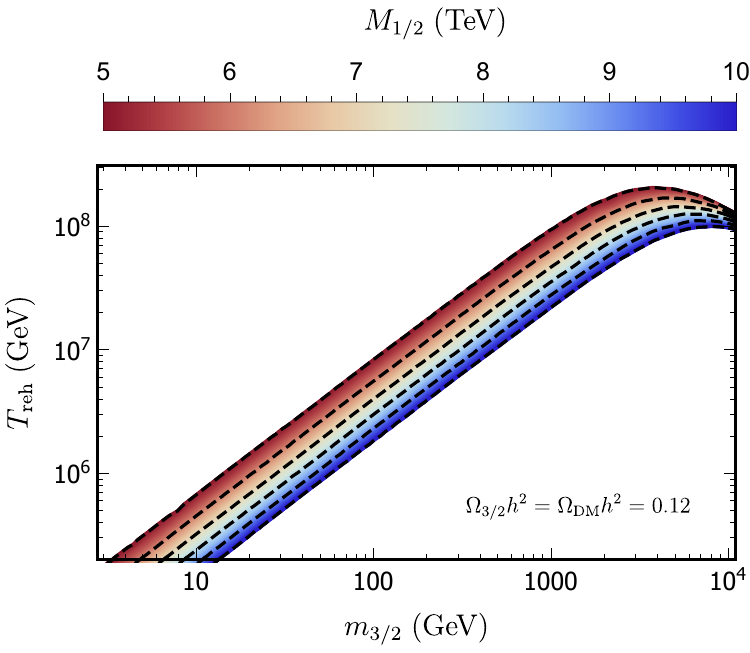} \\[0.2cm]
\includegraphics[width=0.485\textwidth]{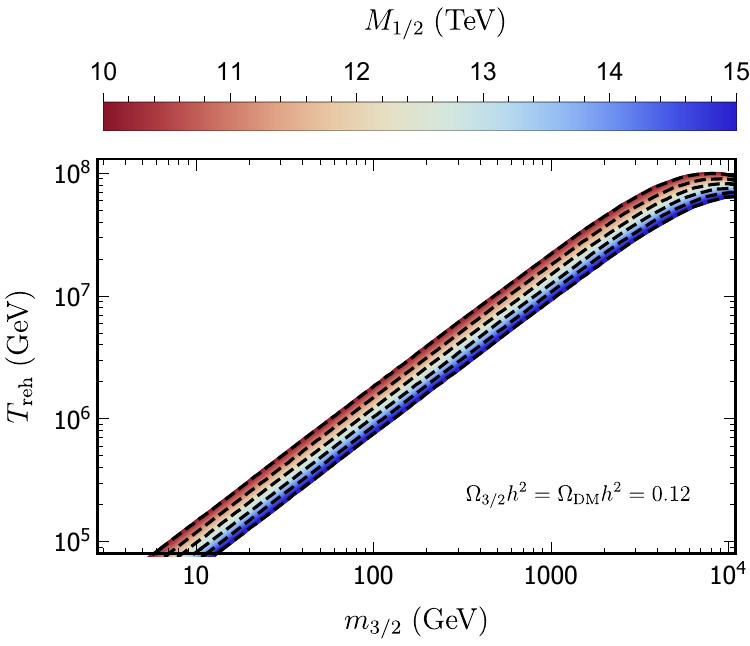} \,
\includegraphics[width=0.485\textwidth]{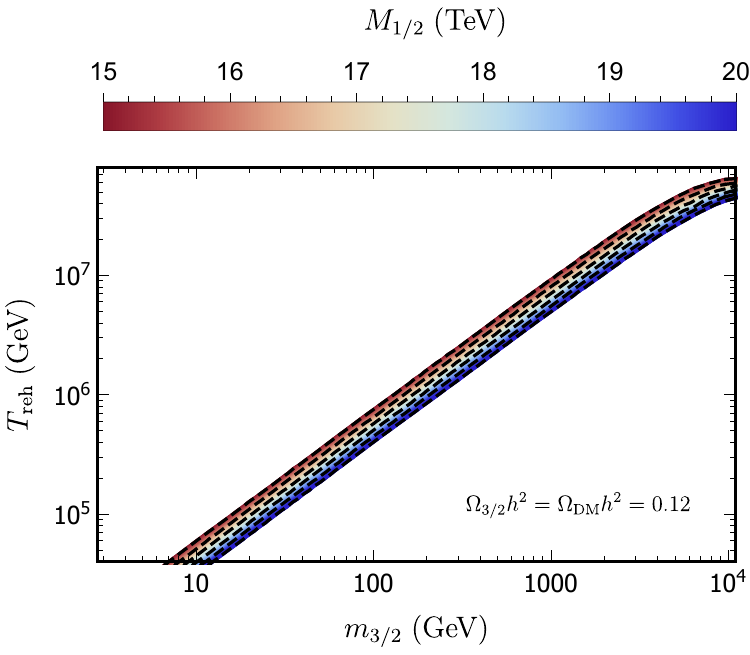}
}
\caption{
The reheating temperature as a function of gravitino mass for fixed  gravitino thermal abundance $\Omega _{3/2}h^2 = \Omega_{\rm DM}h^2 =0.12$.
 The dashed curves, correspond to the values of $M_{1/2}$: (a) from   $ 750\GeV $ to $5\TeV $ (upper left panel),
(b) from   $ 5\TeV $ to $10 \TeV $ (upper right  panel),
(c) from   $ 10\TeV $ to $15 \TeV $ (lower  left  panel),
and (d) from   $ 15\TeV $ to $20 \TeV $ (lower  right  panel), as  
 described in the text. 
 The trilinear coupling $A_t$ is neglected, and the top Yukawa coupling is $y_t = 0.7$.}
\label{fig:Treh_m}
\end{figure}

Before presenting our numerical results, it is helpful to first discuss the various 
DM candidates in SUGRA and how their mass hierarchy can ultimately impact the reheating temperature.
In   extensively studied models like   mSUGRA and   CMSSM, 
the neutralino is the primary DM  candidate, aside from the gravitino.
For example, in the CMSSM framework, where supersymmetry breaking is gravity 
mediated, gauge 
couplings and soft masses are unified at the GUT 
scale, and radiative electroweak symmetry breaking occurs,
the neutralino mass  $m_\chi$  is typically up to $\sim 1/2$ of the common 
gaugino mass  $M_{1/2}$. 
The neutralino is a mixture of bino ($\tilde{B}$), neutral wino ($\tilde{W^3}$), 
and the neutral Higgsinos ($\tilde{H_1^0}$) and ($\tilde{H_2^0}$). 
In particular, if the neutralino is predominantly bino, then  
$m_\chi \simeq 0.45 \, M_{1/2}$. However, if it is mostly Higgsino, 
its mass is closer to  $m_\chi \simeq 0.3 \, M_{1/2}$.

In figure~\ref{fig:Treh_m} 
we plot the reheating temperature as a function of the gravitino mass, having fixed the 
gravitino abundance to $\Omega_{3/2}h^2=0.12$, 
for four ranges of $M_{ 1/2}$: (a) from  $750 \GeV$ to $5\TeV$,
(b) from  $5 \TeV$ to $10\TeV$, (c) from  $10 \TeV$ to $15\TeV$ and
(d) from  $15 \TeV$ to $20\TeV$.
 In the calculation of the gauge couplings and gaugino masses involved in equation~\eqref{eq:dmdensity}, we assume gauge coupling unification and a universal gaugino mass  $M_{1/2}$ at the GUT scale $\sim 2\times 10^{16} \GeV$.
 Moreover, in equation~\eqref{gamma:top} the trilinear coupling $A_t$ is neglected 
 and the top Yukawa coupling is $y_t = 0.7$. 
 The dashed curves in the upper left plot,   figure~\ref{fig:Treh_m} (a), correspond to the values,
$M_{1/2} =750\GeV,  1, 2, 3, 4$ and $5 \TeV $, respectively.  In the subsequent plots, we have included a contour for every $1\TeV$.

Based on  figure~\ref{fig:Treh_m} (a), 
a few  comments on the dependence of $T_{\rm reh}$ as a 
function of $m_{3/2}$ for various values of $M_{1/2}$ and fixed $\Omega_{3/2}h^2$,   are in order.
First, for  large gravitino masses $m_{3/2}\simeq 10^4 \GeV$ all these curves converge  to  ``the point" 
$T_{\rm reh}\simeq 10^8 \GeV$.  This is because,  for such large   $m_{3/2}$   
and fixed $\Omega_{3/2}h^2$, 
the reheating temperature is almost independent of $M_{1/2}$, 
as the factor $M_N^2 / (3 m^2_{3/2})$ in equation~\eqref{eq:dmdensity} becomes negligible.
Second, we observe that  for $M_{1/2} \leq 3 \TeV$ and  $m_{3/2} \leq 1 \TeV$,
the term $M_N^2 / (3 m^2_{3/2})$ in equation~\eqref{eq:dmdensity} leads to an    
increasing   $T_{\rm reh}$ as a function of $m_{3/2}$. However, 
for  $m_{3/2} > 1 \TeV$ this term is subdominant, 
 causing $T_{\rm reh}$ to eventually decrease. 
 For this reason the curves appear to have 
a maximum for the values of $M_{1/2}$ we are plotting. Eventually,
for higher  values of $M_{1/2}$ the $M_N^2 / (3 m^2_{3/2})$ dominates over unity 
in equation~\eqref{eq:dmdensity},  causing the maximum to disappear 
and leaving only a behavior where  $T_{\rm reh}$ is almost proportional to  $m_{3/2}$.
This can be seen in  figure~\ref{fig:Treh_m}~(b), (c) and (d). 

\begin{figure}[t!] 
\centering
{\includegraphics[width=0.485\textwidth]{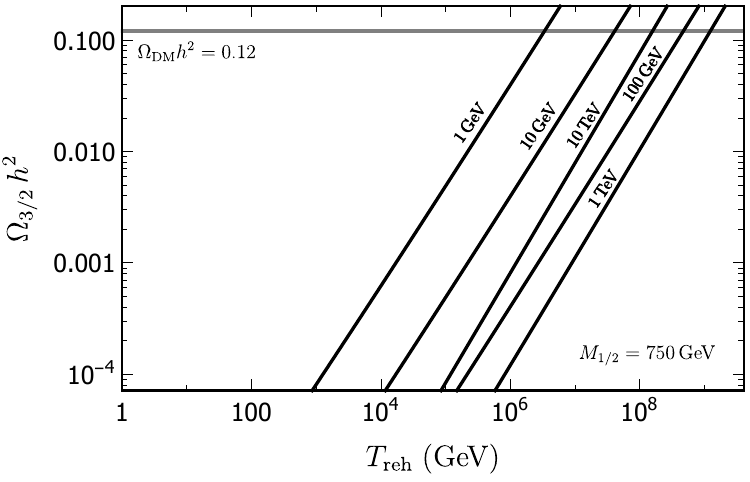} \,
\includegraphics[width=0.485\textwidth]{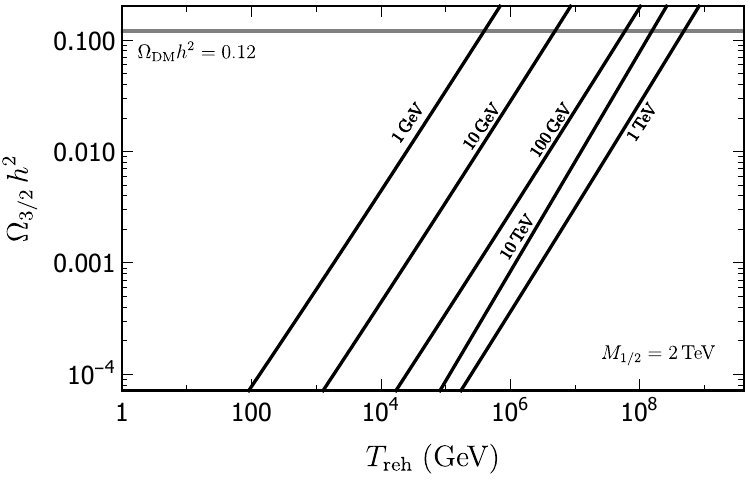} \\[0.2cm]
\includegraphics[width=0.485\textwidth]{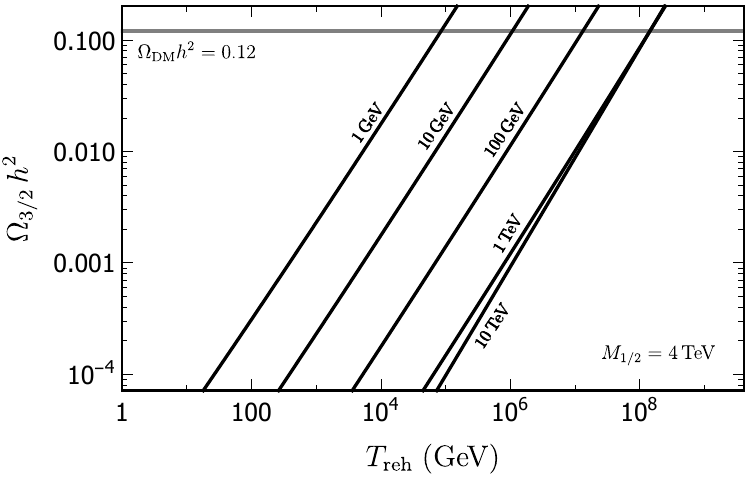} \,
\includegraphics[width=0.485\textwidth]{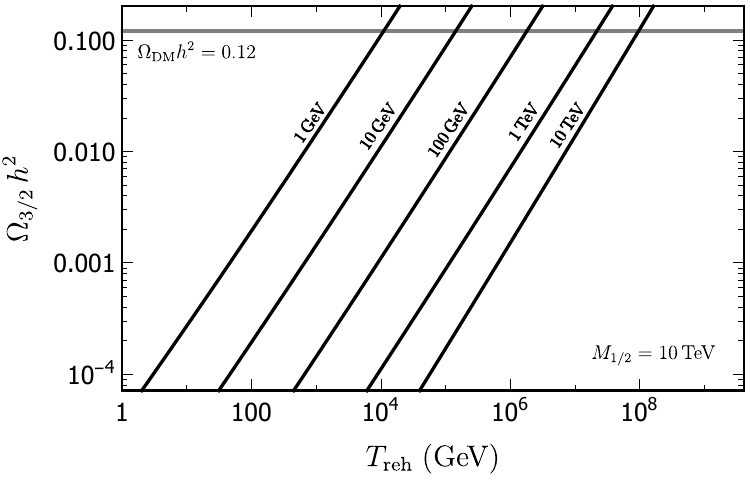}\\[0.2cm]
\includegraphics[width=0.485\textwidth]{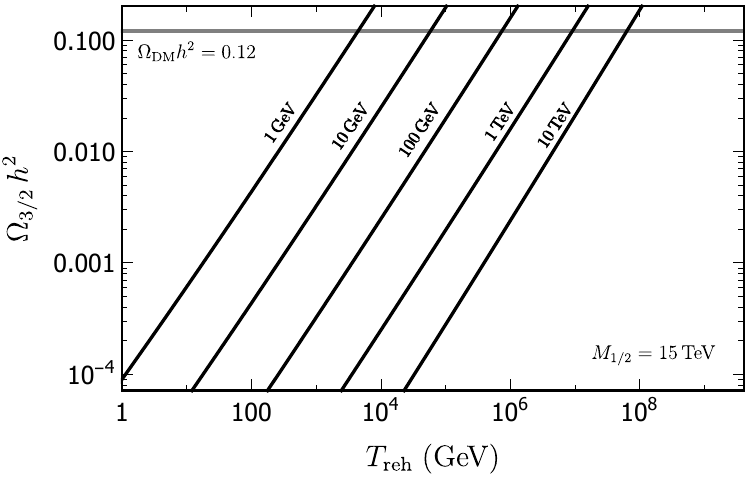} \,
\includegraphics[width=0.485\textwidth]{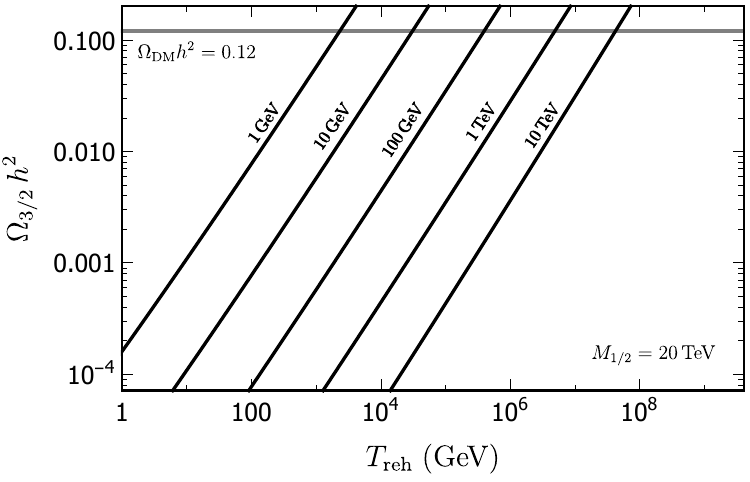} 
}
\caption{The thermal gravitino density, $\Omega_{3/2} h^2$, is shown as a function of the reheating temperature, $T_{\rm reh}$, for various gravitino masses ranging from $1\GeV$ to $10\TeV$. The universal gaugino mass, $M_{1/2}$, varies from $750\GeV$ (top left) to $20\TeV$ (bottom right). The gray horizontal 
thick line  represents the  cosmologically 
allowed DM  density, $\Omega_{3/2}h^2= \Omega_{\rm DM}h^2=0.12 \pm  0.0012$.}
\label{fig:gravdens}
\end{figure}

Moreover, assuming  that $M_{1/2} = 750 \GeV$, as indicated by recent LHC data~\cite{ATLAS:2017weo, CMS:2019zmd} on gluino searches, we can deduce from figure~\ref{fig:Treh_m}~(a) that for a maximum $T_\mathrm{reh} \simeq 10^9  \GeV$, the corresponding gravitino mass is $m_{3/2} \simeq 550 \GeV$. 
On the other hand, taking $M_{1/2}=  5  \TeV$ the maximum reheating temperature is reduced by almost an order 
of magnitude to $T_\mathrm{reh} \simeq 2\times 10^8  \GeV$. However, this temperature 
corresponds to much heavier gravitino $m_{3/2} \simeq 3  \TeV$.

As shown in figure~\ref{fig:Treh_m}~(a) to (d), higher values of $M_{1/2}$ for the same 
gravitino mass result in lower reheating 
temperatures ($T_{\rm reh}$). For instance, with $m_{3/2} = 1 \, \TeV$, 
we find that $T_{\rm reh}$ is greater than $10^8  \GeV$ for $M_{1/2}$ up to 5 TeV, 
whereas $T_{\rm reh}$ falls within the range of $5 \times 10^6  \GeV < T_{\rm reh} < 8 \times 10^6  \GeV$ for $15  \TeV < M_{1/2} < 20  \TeV$.
In addition, we would like to emphasize that although our 
analysis does not assume a specific supersymmetry-breaking scenario, 
such as gravity mediation in the CMSSM, adopting such a framework would 
impact the results. Specifically, there are regions in each plot 
in figure~\ref{fig:Treh_m} where the lightest neutralino is lighter
than the gravitino. For instance, in panel (a), with $M_{1/2} = 1  \TeV$ and assuming 
$m_\chi \simeq 500  \GeV$, the corresponding dashed curve above this value does not correspond to a gravitino dark matter scenario in the CMSSM. These ``neutralino" dark matter regions shrink as $M_{1/2}$ increases in the subsequent panels. Nevertheless, since we have not committed ourselves to this particular SUSY model, our discussion remains general, even though these regions are considered valid.

In figure~\ref{fig:gravdens} we display the 
 thermal gravitino density $\Omega_{3/2} h^2 $  as a function of the reheating temperature $T_{\rm reh}$ 
 for five  values of the gravitino mass: $1, 10, 100 \GeV$ and $1, 10 \TeV$, corresponding
 to the  slanted lines  in each plot.
We present   six  plots for  $M_{1/2}=750\GeV, 2, 4, 10, 15, 20 \TeV$, arranged from
the top left window to the
bottom right, respectively. 
Such high values of $M_{1/2}$ are favoured in regions of the 
parameter space of CMSSM~\cite{Ellis:2022emx} or its extensions~\cite{Ellis:2024ijt}, 
where all the current essential  direct and indirect constraints  are satisfied. 
Assuming that the entire  DM of the Universe is due to thermally 
produced gravitinos, we have marked the  $3\sigma$ cosmologically  allowed region    by the   gray horizontal  thick line. 
However, we also show values   of $\Omega_{3/2} h^2 $ much lower from this value,
under the assumption that the gravitino is not the main DM particle. 
For instance, as discussed in the Introduction, 
this scenario can be realised    in SUGRA-based models, where the neutralino is 
the LSP of the model.

The discussion we have had for figure~\ref{fig:Treh_m}  will help us understand the relative positions of the 
  lines in the plots in figure~\ref{fig:gravdens}. Let’s start with the first plot for
$M_{1/2}=750\GeV$. Assuming   $\Omega_{3/2}= \Omega_{\rm DM}$,   we get $T_{\rm reh}\simeq 10^9\GeV$ 
for $m_{3/2}=1\TeV$, as   expected from  figure~\ref{fig:Treh_m}~(a). 
Relaxing the DM constraint in  this case, we notice that the 
reheating temperature can be reduced by at least three order of magnitudes to 
$T_{\rm reh}\lesssim  10^6\GeV$. The same applies to all $m_{3/2}, M_{1/2}$ values (slanted lines) in these plots.

However in the first  plot, 
it is interesting to note that the line for $m_{3/2}=10\TeV$ is to the left of the
lines for $m_{3/2}=100\GeV$ and   $m_{3/2}=1\TeV$.  This again results from the maximum  observed in figure~\ref{fig:Treh_m}~(a), as discussed above, which leads to a smaller $T_{\rm reh}$ for $m_{3/2}=10\TeV$ compared to  $m_{3/2}=100\GeV$. The same pattern is observed in the plot for $M_{1/2}=2 \TeV$. In the 
plot for  $M_{1/2}=4 \TeV$,  the lines for  $m_{3/2}=1\TeV$ and $m_{3/2}=10\TeV$ almost coincide.
The last three plots of figure~\ref{fig:gravdens},   
 where $m_{3/2}> 5\TeV$,  the expected order of the lines appears. 
These correspond to   figure~\ref{fig:Treh_m}~(b), (c) and (d), 
where the maximum has almost disappeared.

As anticipated, if the entire  DM content is not attributed solely to the thermal abundance of gravitinos, the upper bounds on the reheating temperature are significantly lowered. This reduction can be achieved through various methods. One approach involves considering alternative DM candidates, such as axions or axinos. Specifically, within the frameworks of minimal SUGRA model or CMSSM, if the gravitino is the LSP, it can also be generated through the late decays of NLSP, like gauginos (neutralino)  or sfermions. Consequently, the thermal abundance of gravitinos may only contribute a portion of the total DM density in the Universe. Similar to figure~\ref{fig:Treh_m}, figure~\ref{fig:gravdens} also features lines that, within the context of the CMSSM, correspond to a neutralino DM  scenario. For instance, in the first plot, assuming   $m_\chi \simeq 500 \GeV$  the lines corresponding to 
 $m_{3/2}= 1\TeV, 10\TeV$  fall within this category. 
As discussed in the introduction, since our model is SUGRA-based, 
gravitinos continue to be produced thermally but eventually decay with a 
width given by equation~\eqref{eq:chi_decay}, contributing to the total neutralino 
DM through equation~\eqref{eq:extra_relic2}. 
Consequently, in this scenario, the   $T_{\rm reh}$
inferred from figure~\ref{fig:gravdens} is significantly
lower than what would be required if $\Omega_{\rm DM}h^2 = 0.12$.

\section{Conclusions}
\label{sec:concl}
In this paper, we have calculated the gravitino production rate and its
thermal abundance,  using the complete one-loop thermally corrected gravitino self-energy. 
We have refined the main analytical formulas for the gravitino decay rate and computed it numerically without approximation, appropriately parameterizing the final results. 
Within the framework of minimal supergravity models, assuming gaugino mass 
unification, we have updated the bounds on the reheating temperature for 
specific gravitino masses.

Assuming that gravitinos constitute the entirety of the dark matter in the Universe,  $\Omega_{3/2}h^2 = \Omega_{\rm DM}h^2 \sim 0.12$, imposes stringent constraints on the reheating temperature.  For instance, 
with a gluino mass at the current LHC limit, we find that a maximum reheating temperature of $T_\mathrm{reh} \simeq 10^9 \, \mathrm{GeV}$ is compatible with a gravitino mass of $m_{3/2} \simeq 1 \, \TeV$.
Moreover, with a reheating temperature  an order of magnitude lower,
$T_\mathrm{reh} \simeq 10^8 \, \mathrm{GeV}$, the common gaugino mass $M_{1/2}$ can range up to 2-4 TeV within the same gravitino mass range.

For much higher values of  $M_{1/2}$  which are favored by current accelerator and cosmological data
 in the context of supersymmetric models, e.g., 
$M_{1/2}=10 \TeV$,  and $m_{3/2} \simeq 1 \, \TeV$, the reheating temperature 
that is compatible with the gravitino DM scenario,
is   $T_\mathrm{reh} \simeq 10^7$. 
Allowing for the possibility that other particles, 
such as the neutralino, can play the role of dark matter, 
we can relax the gravitino DM constraint. 
In this case the reheating temperature can be much smaller.
Specifically,   for $m_{3/2} \simeq 1 \, \TeV$ and $M_{1/2}=10\TeV$ ($M_{1/2}=20\TeV$)
we obtain an upper bound for the reheating temperature $T_{\rm reh} \lesssim 10^4 \TeV$
($T_{\rm reh} \lesssim 10^3 \TeV$), respectively.

It should be noted that applying constraints on gravitino dark matter scenarios using $T_\mathrm{reh} $ 
implies the assumption of thermal leptogenesis as the mechanism for generating baryon asymmetry. However, many alternative models for baryogenesis exist. Additionally, as previously highlighted, the thermal gravitino abundance is only a part of the total abundance, which impacts the phenomenological analysis.
Moreover, our results for the gravitino decay rate and its abundance can be applied within a general supersymmetric framework, regardless of the specifics of the supersymmetry-breaking scenario

\vspace{1cm}
%-------------------------------------------------------------------------------
\acknowledgments
%-------------------------------------------------------------------------------
The work of IDG was supported by the Estonian Research Council grants MOB3JD1202, RVTT3,  RVTT7, and by the CoE program TK202 ``Fundamental Universe''. 
The  work of V.C.S. was partially supported by the Hellenic Foundation for Research and Innovation (H.F.R.I.) under the ``First Call for H.F.R.I. Research Projects to support Faculty members and Researchers and the procurement of high-cost research equipment grant'' (Project Number: 824). V.C.S.  is  grateful to the William I. Fine Theoretical Physics Institute at the University of Minnesota for their financial support 
and the warm hospitality extended to him  during
his sabbatical leave. He also   acknowledges support from the Simons Foundation Targeted Grant 920184 to the William I. Fine Theoretical Physics Institute.

\newpage 
\appendix
\section{Amplitudes}
\label{appendix:A}

\subsection{Couplings}
\label{appendix:langr}
The involved interactions to the gravitino $\Gr$ can be found 
e.g. in ~\cite{Eberl:2015dia}. The SM and MSSM interactions are given in corresponding
textbooks. But for completeness and simplicity we list all involved interactions below.\\

\noindent
The $g g \sg \Gr$ interaction is
\begin{eqnarray}
\label{lag3color1}
& {\cal L} = & - {i \over 8 \mplanck} \left(2 \partial_\r G^a_\s -
g_3 f^{a b c} G^b_\r G^c_\s \right) \left(\overline\grav_\m [\g^\r,\g^\s] \g^\m  \tilde g^a + \overline{\tilde g^a} \g^\m [\g^\r,\g^\s] \grav_\m\right)\, .
\end{eqnarray}
Note that the vertex rule of $g_\mu^a g_\nu^b \sg^c \Gr_\rho$ gets an additional factor of 2,
\begin{equation}
{\rm Vertex}(g_\mu^a g_\nu^b \sg^c \Gr_\rho) =  - {i g_3 \over 4 \mplanck}  f^{a b c} \g^\r  [\g^\m,\g^\n]\, .
\end{equation}
The vertex rule for $g_\mu^a g_\nu^b \Gr_\rho$ can also be deduced from (\ref{lag3color1}), with $i \partial_\m \to p_\m$ (incoming).\\
The three-gluon vertex, defined by the Lagrangian ${\cal L} = - g_3  f^{a b c} G^{b \n} G^{c \m} \left(\partial_\nu G_\mu^a \right)$ is
\begin{equation}
{\rm Vertex}(g_\mu^a g_\nu^b g^c_\rho) =  - {g_3 \over 4 \mplanck}  f^{a b c} V^{\mu\nu\rho}(k_1, k_2, k_3) \, ,
\end{equation}
with all momenta defined incoming, and
\begin{equation}
V^{\mu\nu\rho}(k_1, k_2, k_3)  = (k_1 - k_2)^\r g^{\m\n} + (k_2 - k_3)^\m g^{\n\r} +  (k_3 - k_1)^\n g^{\m\r}\, . 
\end{equation}
For the $\xi = 1$ gauge we also will need the coupling of the FP-ghosts to the gluon. The Lagrangian is
\begin{equation}
{\cal L} = - g_3 f_{abc} (\partial \bar\eta^a) G^{c \mu} \eta^b\,,
\end{equation}
with $\bar\eta = \eta^*$.  Note that the $\eta$ and $\bar\eta$ are anti-commuting scalar fields. The gluon-$\eta$-$\bar\eta$ vertex is
\begin{equation}
{\rm Vertex}(g_\mu^c \to \eta^a(k_2) \bar\eta^b) =  - g_3 f_{abc} k_{2 \mu}\, .
\end{equation}
Next we have the $g \sg \sg$ interaction. The Lagrangian is
\begin{equation}
 {\cal L} = {i \over 2} g_3 f_{a b c} G^a_\m \bar\sg^b \g^\m \sg^c 
\end{equation}
with e. g. 
\begin{equation}
{\rm Vertex}(g_\mu^a \sg_\nu^b \sg^c) = - g_3  f_{a b c} \e_\mu (k_0) \bar u(k_1) \g^\m v(k_2)\, , 
\end{equation}
taking $\sg(k_1)$ as outgoing particle and $\sg(k_2)$ as outgoing antiparticle. Note the additional factor 2 due to the Majorana structure of $\sg$.

Now we turn  to systems involving strong interacting quarks and squarks.
We will use the shortcuts
\begin{eqnarray}
\a_{RL}^i & = & R_{iL} P_R - R_{iR} P_L\, ,\\
\a_{LR}^i & = & R_{iL} P_L - R_{iR} P_R\, ,
\end{eqnarray}
and define the conjugated expressions as
\begin{eqnarray}
\a_{RL}^{i*} & \equiv & R^*_{iL} P_R - R^*_{iR} P_L\, ,\\
\a_{LR}^{i*} & \equiv & R^*_{iL} P_L - R^*_{iR} P_R\, ,
\end{eqnarray}
in the convention $\sq_i = R_{i\alpha}, \, i = 1,2; \alpha = L,R$.
For massless squarks $\sq_1 = \sq_L$ and $\sq_2 = \sq_R$,
$R_{1L} = R_{2R} = 1, R_{1R} = R_{2L} = 0$.\\
The $q \sq \sg$ interaction is given by
\begin{equation}
 {\cal L} = - \sqrt{2} g_3 T^a_{s t} \left( \bar q^s \a_{RL}^{i*} \sg^a \sq_i^t + \bar \sg^a \a_{LR}^i  q^t \sq^{*\, s}_i\right)\,,
\end{equation}
and $q \sq \Gr$ by
\begin{equation}
 {\cal L} =  {1 \over \sqrt2 \mplanck} \delta_{r t}\left[ \bar q^r \g^\n \g^\m \a_{RL}^{i*}  \Gr_\n  (i \partial_\mu \sq^t_i)
 - \Grb \g^\m \g^\n \a_{LR}^i  q^t  (i \partial_\mu \sq_i^{* r})\right]\, . 
\end{equation}
We also need the $q q g$ and $\sq \sq g$ interactions,
\begin{equation}
 {\cal L} = - g_3 T^a_{s t} G_\m^a \bar q^s \g^\mu q^t\, ,
\end{equation}
\begin{equation}
 {\cal L} = - g_3 T^a_{s t} G_\m^a  \sq_i^{* s}\Big[(\partial^\m \sq_j) - (\partial^\m  \sq_i^{* s}) \sq_j)\Big] \delta_{ij}\, ,
\end{equation}
with e.g. the rule for
\begin{equation}
{\rm Vertex}\big( \sq^r_i(k_1) \to g^a + \sq^s_j(k_2)\big) = - i g_3 T^a_{rs} (k_1 + k_2)_\m \delta_{ij} \, .
\end{equation}

\subsection{Kinematics}
\label{kinematics}
 
 Before we go into the details of the individual channels we fix the kinematics by
 \begin{equation}
P_1 (k_1) +  P_2 (k_2) \to P_3 (p_1)  + \Gr (p_2) \, ,
\end{equation}
with the particles $P_{1,2,3}$ and the overall momenta relation $k_1 + k_2 = p_1 + p_2$.
The Mandelstam variables are fixed by
\begin{eqnarray}
s & = &  (k_1 + k_2)^2 = (p_1 + p_2)^2 \,, \\
t & = & (k_1 - p_1)^2 = (k_2 - p_2)^2\,, \\
u & = & (k_1 - p_2)^2 = (k_2 - p_1)^2\,,
\end{eqnarray}
together with $s + t + u = \sum_{i=1}^4  M_i^2$.
In general, we can have three types of crossing symmetries in the amplitudes,
\begin{eqnarray}
\label{P1toP2}
P_1 \leftrightarrow P_2: & k_1 \leftrightarrow k_2 \quad  \ldots\,  & s {\rm ~ unchanged}\,,  t \leftrightarrow  u\,, \\
\label{P1toP3}
P_1 \leftrightarrow P_3: & k_1 \leftrightarrow p_1 \quad   \ldots\,  & t {\rm ~ unchanged}\,,   s \leftrightarrow - u\,, \\
\label{P2toP3}
P_2 \leftrightarrow P_3: & k_2 \leftrightarrow  p_1 \quad   \ldots\, & u {\rm ~ unchanged}\,,  s \leftrightarrow - t\, ,
\end{eqnarray}
 assuming massless particles.
 In the cases where crossing symmetries are possible, we will discuss them in the following sections.
 Furthermore the case (\ref{P1toP2}) is of no relevance because interchanging $P_1$ with $P_2$ will give the same 
 cross section.  Note that for (\ref{P1toP3}) and (\ref{P2toP3}) we have particle $\leftrightarrow$ antiparticle.
 The gluon $g$ and the gluino $\sg$, which is a Majorana particle, remain unchanged.\\

\subsection{Involved interactions}
In what follows we present the $|{\cal M}_{X,\rm full}|^2$ for the above processes as it is also shown in table~\ref{table:1}.
We begin by focusing exclusively on the $\suthree$ results (see figures.~\ref{process_F} - \ref{process_G}). All results for  $|{\cal M}_i|^2 (i = A, \ldots J)$
are presented in terms of  $g_3^2  Y_3 / \mplanck^2$ , with  $Y_3$  defined in~\eqref{eq:yfactor}.\\

\noindent
\textbf{$\bullet $ Amplitudes for $\sg \sg \to \sg \Gr$ (process F) }
 
 \begin{figure}[h] % H % H  
\centering
\includegraphics[width=\textwidth]{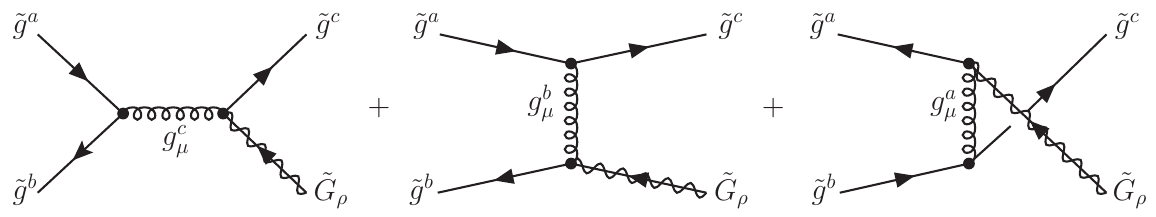}
\caption[]{Feynman graphs for the process  $\sg \sg \to \sg \Gr$. 
\label{process_F}}
\end{figure}
There are three Feynman diagrams,  with a $g$ propagator in the $s$-channel (1),
$t$-channel (2) and $u$-channel (3). Here
one has to be careful because ${\cal M}_2$ gets an additional minus sign from fermion statistics.
\begin{eqnarray}
 {\cal M}_{1} & = & - {g_3 \over 4  \mplanck s} f_{a b c} \bar v(k_2) \g_\m u(k_1) \bar u(p_1) \g^\r [\slash{k}_1 + \slash{k}_2, \g^\m] v_\rho(p_2)\,, \\
 {\cal M}_{2} & = &  {g_3 \over 4  \mplanck t } f_{a b c}  \bar u(p_1) \g_\m u(k_1)  \bar v(k_2) \g^\r [\slash{p}_1 - \slash{k}_1, \g^\m] v_\rho(p_2)\,, \\ 
 {\cal M}_{3} & = &  -{g_3 \over 4  \mplanck u} f_{a b c}  \bar u(p_1) \g_\m u(k_2)  \bar v(k_1) \g^\r [\slash{k}_2 - \slash{p}_1, \g^\m] v_\rho(p_2)\, .
 \end{eqnarray} 
The full amplitude squared, with fixed $a, b, c$, is
\begin{equation}
|{\cal M}_{F,\rm full}|^2 =- 8 {(s^2 + s t + t^2)^2 \over s t (s + t)} |f^{a b c}|^2 \to 8 C_3 {(s^2 + t^2 + u^2)^2 \over s t u}  \, ,
\end{equation}
where we used that $s + t = - u$ and $|f^{a b c}|^2  \to C_3 =24 $ for $\suthree$. The factor of $1/2$, accounting for the identical incoming gluinos, is already included.
Note that there is no crossing channel possible, because we get for the interchanges~(\ref{P1toP2}) -  (\ref{P2toP3}) always the same sum of amplitudes.\\

\noindent
\textbf{$\bullet$ Amplitudes for $g g \to \sg \Gr$ (process A)}\\
\begin{figure}[h]
\centering
\includegraphics[width=0.67\textwidth]{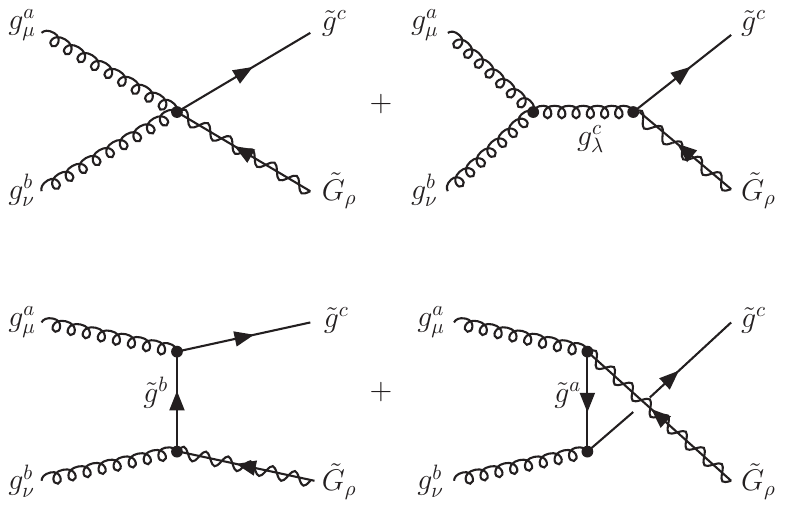}
\caption[]{Feynman graphs for the process  $g g \to \sg \Gr$ . 
\label{process_A}}
\end{figure}
There are four Feynman diagrams, (1): with the four-point interaction, (2): with a $g$ propagator in the $s$-channel,
(3): a  $\sg$ propagator in the $t$-channel, and (4): a  $\sg$ propagator in the $u$-channel,
\begin{align}
 {\cal M}_{1} & =  - {g_3 \over 4  \mplanck} f_{a b c} \bar u(p_1) \g^\r [ \g^\m, \g^\n] v_\rho(p_2) \e_\mu(k_1) \e_\n(k_2)\, . \\
 {\cal M}_2 & =  {g_3 \over 4  \mplanck s} f_{a b c} V^{\m\n\d}(k_1, k_2,\! - k_1\!-\!k_2) 
\bar u(p_1) \g^\r [\slash{k}_1 + \slash{k}_2, \g_\d] v_\rho(p_2) \e_\mu(k_1) \e_\n(k_2)\, ,\\
 {\cal M}_3 & =  {g_3 \over 4  \mplanck} {f_{a b c} \over t - m_\sg^2} \bar u(p_1) \g^\m (\slash{k}_2 - \slash{p}_2 + m_\sg) \g^\r [\slash{k}_2, \g^\n] v_\rho(p_2) 
 \e_\mu(k_1) \e_\n(k_2)\, ,\\
 {\cal M}_4 & =  -{g_3 \over 4  \mplanck} {f_{a b c} \over u - m_\sg^2} \bar u(p_1) \g^\n (\slash{k}_1 - \slash{p}_2 + m_\sg) \g^\r [\slash{k}_1, \g^\m] v_\rho(p_2) 
 \e_\mu(k_1) \e_\n(k_2)\, .
 \end{align}
  \begin{figure}[h] % H  
\centering
\includegraphics[width=0.67\textwidth]{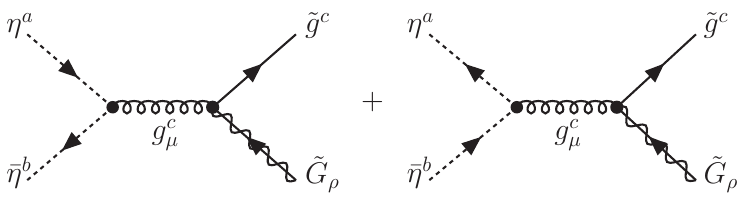}
\caption[]{Feynman graph with FP-ghosts in the $\xi = 1$ gauge for the process A. 
\label{process_A_ghosts}}
\end{figure}
 For the $\xi = 1$ gauge we also need the matrix elements with the incoming FP-ghosts for the gluon. There are two graphs possible,
\begin{eqnarray}
 {\cal M}_{\eta} & = &  {g_3 \over 4  \mplanck} f_{a b c} \bar u(p_1) \g^\r [\slash{k}_1 +  \slash{k}_2, \slash{k}_2] v_\rho(p_2) \, , \\ 
 {\cal M}_{\bar\eta} & = & - {g_3 \over 4  \mplanck} f_{a b c} \bar u(p_1) \g^\r [\slash{k}_1 +  \slash{k}_2, \slash{k}_1] v_\rho(p_2) \, .
 \end{eqnarray} 
The result for the squared amplitudes for the massless case is
\begin{eqnarray}
|{\cal M}_{\eta}|^2 +     |{\cal M}_{\bar\eta}|^2  = - {g_3^2 C_3 \over 2  \mplanck s} (s + 2 t)^2\, ,
\end{eqnarray}
with $|f^{a b c}|^2  \to C_3$.
The full amplitude squared for fixed $a, b, c$ is
\begin{equation}
|{\cal M}_{A,\rm full}|^2 = 2 \left( s + 2 t + 2 {t^2 \over s} \right) |f^{a b c}|^2 \to 2C_3 \left( s + 2 t + 2 {\frac{t^2}{s}} \right)  \, ,
\end{equation}
where the factor of $1/2$  (which accounts for the identical incoming gluons) is already included in the result. We also use the fact that  $|f^{a b c}|^2  \to C_3 $.
The amplitudes for $g \sg \to g \Gr$ named process B we get from A using~(\ref{P2toP3}).
\\

\noindent
\textbf{$\bullet $ Amplitudes for $\sq_i \sg \to \sq_j \Gr$ (process H)}\\
\begin{figure}[h] % H  
\centering
\includegraphics[width=\textwidth]{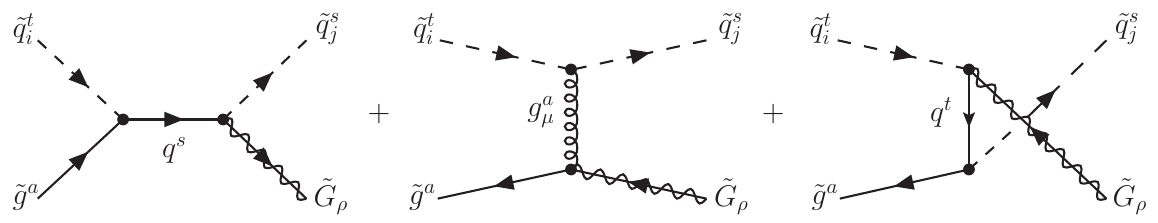}
\caption[]{Feynman graphs for the process  $\sq_i \sg \to \sq_j \Gr$ . 
\label{process_H}}
\end{figure} 
There are three Feynman diagrams, (1): with a $q$ propagator in the $s$-channel, (2): with a $g$ propagator in the $t$-channel and
(3): a  $q$ propagator in the $u$-channel.
Here one has to be careful again because ${\cal M}_1$ gets an additional minus sign due to fermion statistics.
\begin{eqnarray}
{\cal M}_{1} & = & - {i g_3 \over \mplanck} {T^a_{st}  \over s - m^2_q}  \bar u_\r(p_2) \slash{p}_1 \g^\r  \a_{LR}^j  (\slash{k}_1 + \slash{k}_2 + m_q) \a_{RL}^{i*} u(k_2)\, , \\
{\cal M}_{2} & = & {i g_3 \over 4 \mplanck t} T^a_{st} \bar v(k_2) \g^\r [\slash{k}_1 - \slash{p}_1,  \slash{k}_1 + \slash{p}_1] v_\r(p_2) \d_{i j}\, , \\
{\cal M}_{3} & = &  {i g_3 \over \mplanck} {T^a_{st}  \over u - m^2_q}   \bar v(k_2)  \a_{LR}^j (\slash{k}_1 - \slash{p}_2 + m_q) \g^\r \slash{k}_1 \a_{RL}^{i*} v_\r(p_2) \, .
\end{eqnarray}
The full amplitude squared, with fixed color indices and fixed $i,j$ is
\begin{equation}
\label{eq:2to2H}
|{\cal M}_{H,\rm full}|^2 =- 2 \left(t + 2 s + {s^2 \over t} \right)  |T^a_{r s}|^2 \delta_{i j} \to - 4 C'_3\left(t + 2 s + {s^2 \over t} \right)\,.
\end{equation}
Now because we consider 6 squarks--anti-squarks pairs we get a factor of 2, for the two mass eigenstates
for each flavour.
 Thus for this process  we get  $|T^a_{r s}|^2 \to 2 C'_3$.
The amplitudes for $\sq_i \sq^*_j \to \sg \Gr$ named process J we get from H using~\eqref{P2toP3}.\\

\noindent
\textbf{$\bullet $ Amplitudes for $\sq_i g \to q \Gr$ (process C)}\\
\begin{figure}[h] % H  
\centering
\includegraphics[width=0.67\textwidth]{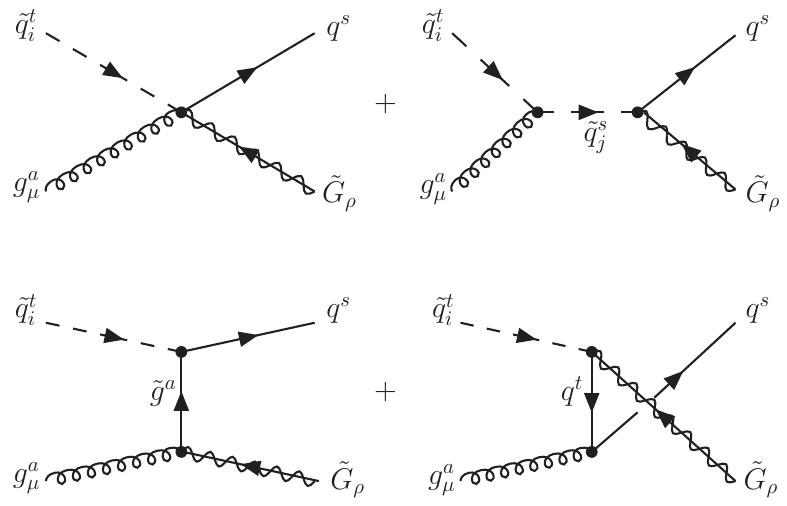}
\caption[]{Feynman graphs for the process  $\sq_i g \to q \Gr$ . 
\label{process_C}}
\end{figure} 
\noindent
There are four Feynman diagrams, (1): with the four-point interaction, (2): with a $\sq$ propagator in the $s$-channel and
(3): a  $\sg$ propagator in the $t$-channel, and (4): a  $q$ propagator in the $u$-channel,
\begin{eqnarray}
 {\cal M}_1 & = &  - {i g_3 \over \sqrt2  \mplanck} T^a_{st} \bar u(p_1)  \g^\r \g^\m \a_{RL}^{i*}  v_\rho(p_2) \e_\mu(k_2)\, ,\\
 {\cal M}_2 & = &   {i g_3 \over \sqrt2  \mplanck} {T^a_{st} \over s - m^2_{\sq_i}} \bar u(p_1)  \g^\r ( \slash{k}_1 + \slash{k}_2) 
 \a_{RL}^{i*}  v_\rho(p_2) (2 k_1 + k_2)^\m  \e_\mu(k_2)\, ,\\
  {\cal M}_3 & = &   {i \sqrt2 g_3 \over 4  \mplanck} {T^a_{st} \over t - M_{N_3}^2} \bar u(p_1) \a_{RL}^{i*} 
 ( \slash{k}_1 - \slash{p}_1 - m_\sg) \g^\r [  \slash{k}_2, \g^\m]  v_\rho(p_2) \e_\mu(k_2)\, ,\\
  {\cal M}_4 & = &   {i g_3 \over \sqrt2  \mplanck} {T^a_{st} \over u - m^2_q} \bar u(p_1)  \g^\m (\slash{k}_1 - \slash{p}_2 + m_q)
  \g^\r   \slash{k}_1  \a_{RL}^{i*}  v_\rho(p_2)  \e_\mu(k_2)\, .
 \end{eqnarray}
The full amplitude squared, with fixed color indices $a,r,s$ and fixed sfermion mass eigenstate indices $i,j$, is
% (= 1 or 2) 
\begin{equation}
|{\cal M}_{C,\rm full}|^2 = 2 s  |T^a_{r s}|^2 \delta_{i j} \to 4s C'_3 \, .
\end{equation}
With 6 squarks and 6 anti-squarks, and $\sum_{a,r,s}  |T^a_{r s}|^2 = 4$, along with the sum over $i, j$  yielding 2, this results in a factor of $2 C'_3$, where $C'_3 = 4 \times 2 \times 6 = 48$.
The amplitudes for $\sq_i \bar q \to g \Gr$ named process E we get from C using (\ref{P2toP3}).
The amplitudes for $g \bar q \to \sq^*_i \Gr$ named process D we get from E using~\eqref{P1toP3}.\\

\noindent
\textbf{$\bullet $ Amplitudes for $q \sg \to q \Gr$ (process G)}\\
\begin{figure}[h] % H  
\centering
\includegraphics[width=\textwidth]{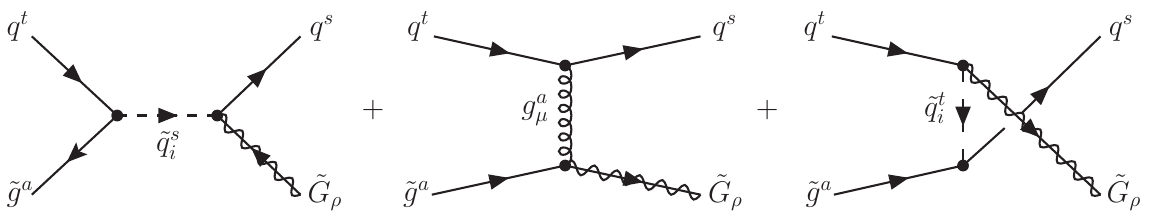}
\caption[]{Feynman graphs for the process  $q \sg \to q \Gr$ . 
\label{process_G}}
\end{figure}  
There are three Feynman diagrams, (1): with a $\sq_i$ propagator in the $s$-channel, (2): with a $g$ propagator in the $t$-channel and
(3): a  $\sq_i$ propagator in the $u$-channel.
Here one has to be careful again because ${\cal M}_3$ gets an additional minus sign from fermion statistics.
\begin{eqnarray}
{\cal M}_{1} & = & \sum_{i = 1}^2 {i g_3 \over \mplanck} {T^a_{st}  \over s - m^2_{\sq_i}} \bar v(k_2) \a_{LR}^i u(k_1)  \bar u(p_1) \g^\m (\slash{k}_1 + \slash{k}_2) \a_{RL}^{i*} v_\m(p_2) \,, \\
{\cal M}_{2} & = & {i g_3 \over 4  \mplanck} {T^a_{st} \over t} \bar u(p_1) \g_\n u(k_1) \bar u_\m(p_2) [ \slash{k}_1 - \slash{p}_1, \g^\n]  \g^\m u(k_2)\, \\
{\cal M}_{3} & = & \sum_{i = 1}^2 {i g_3 \over \mplanck} {T^a_{st}  \over u - m^2_{\sq_i}} \bar u(p_1) \a_{RL}^{i*} u(k_2) \bar u_\m(p_2) (\slash{p}_2 - \slash{k}_1) \g^\m  \a_{LR}^{i*} u(k_1) \,,
\end{eqnarray} 
The full amplitude squared, with fixed $a, r, s$, is
\begin{equation}
|{\cal M}_{G,\rm full}|^2 =- 4 \left(s + {\frac{s^2}{t}} \right)  |T^a_{r s}|^2 \to -8C'_3 \left(s + {s^2 \over t} \right)\, .
\end{equation}
Since there are 6 quarks--anti-quarks pairs, we have $|T^a_{r s}|^2 \to 2 C'_3$.
The amplitudes for $q \bar q \to g  \Gr$ named process I we get from G using(\ref{P2toP3}).\\

The remaining processes can be derived using crossing symmetries, with only 
potential differences in overall factors. For example, process B does not include the additional factor of $ \frac{1}{2} $
that is present in process A. Processes D and E share the same overall  factors as process C. Process I, however, involves an incoming  quark and antiquark, resulting in a factor $2$ less than in process G, where we summed over quarks and antiquarks.
A similar reasoning applies to process J in comparison to process H.

\subsection{Goldstino production}
We have two possibilities to calculate that processes, the so-called derivative and non-derivative approach.
The total amplitudes must be the same in both approaches~\cite{Lee:1998aw}.

\subsubsection{The derivative approach}
\label{sect:deriv}
According to the equivalence theorem in the derivative approach, the gravitino spinors must be replaced in all amplitudes by goldstino spinors using the following rules:
\begin{equation}
v_\mu(p_2) \to i \sqrt{2 \over 3} {1 \over m_{3/2}} v(p_2)\, p_{2 \mu}\, , \quad
\bar u_\mu(p_2) \to i \sqrt{2 \over 3} {1 \over m_{3/2}} \bar u(p_2)\, p_{2 \mu}\, . 
\label{equi_theo}
\end{equation}
 
\subsubsection{The non-derivative approach}
\label{sec:non_der}
Here one has new Feynman rules, which can be found e.g. in the \cite{Bolz:2000xi} or in~\cite{Pradler:2006tpx}.
The goldstino-fermion-sfermion-gluon vertex vanishes and there occurs a new coupling, goldstino-gaugino-sfermion-sfermion.
The goldstino-fermion-sfermion coupling is proportional to $m^2_f - m^2_{\tilde f}$. All the other couplings are 
proportional to the gaugino mass term $M_i$ ~$(M_3 = m_\sg)$.  We are only interested in the massless limit. Thus we set 
the goldstino-fermion-sfermion coupling to zero. We again give only the results for $\suthree$. 
For convenience we  define the  prefactor $P_f$ 
\begin{equation}
P_f \equiv  {m_\sg \over 2 \sqrt{6} \mplanck m_{3/2}} g_3 ( f_{a b c} \,|\, T^a_{rs}) \, ,
\end{equation}
where $f_{a b c}$ ($T^a_{rs}$) is taken in processes without (with) quarks and squarks.
In the amplitudes we set all other occurring $m_\sg$ to zero.

For $\sg \sg \to \sg \grav$, named process F in table~\ref{table:1} we again have
three Feynman diagrams, with a $g$ propagator in the $s$-channel (1),
$t$-channel (2) and $u$-channel (3), all connected by crossing symmetries.  
${\cal M}_2$ gets an additional minus sign from fermion statistics.
\begin{eqnarray}
{\cal M}_{1} & = &  i {P_f \over s}  \bar v(k_2) \g_\m u(k_1) \bar u(p_2) [\slash{k}_1 + \slash{k}_2, \g^\m] v(p_1)\,, \\
{\cal M}_{2} & = &   -i {P_f \over t} \bar u(p_1) \g_\m u(k_1) \bar u(p_2) [\slash{p}_1 - \slash{k}_1, \g^\m] u(k_2)\,, \\
{\cal M}_{3} & = &  i {P_f \over t}  \bar u(p_1) \g_\m u(k_2) \bar u(p_2) [\slash{p}_2 - \slash{k}_1, \g^\m] u(k_1)\,.
\end{eqnarray} 
  
For $g g \to \sg \grav$, named process  A in table~\ref{table:1} we again have
four Feynman diagrams, (1): with the four-point interaction, (2): with a $g$ propagator in the $s$-channel,
(3): a  $\sg$ propagator in the $t$-channel, and (4): a  $\sg$ propagator in the $u$-channel,
\begin{eqnarray}
 {\cal M}_{1} & = & i\, P_f\, \bar u(p_2) [ \g^\m, \g^\n] v(p_1) \e_\mu(k_1) \e_\n(k_2)\, . \\
 {\cal M}_2 & = &- i {P_f  \over s} V^{\m\n\r}(k_1, k_2,\! - k_1\!-\!k_2) 
\bar u(p_2) [\slash{k}_1 + \slash{k}_2, \g_\r] v(p_1) \e_\mu(k_1) \e_\n(k_2)\, ,\\
 {\cal M}_3 & = & i {P_f \over t} \bar u(p_2)  [\slash{k}_2, \g^\n]  (\slash{k}_1 - \slash{p}_1) \g^\m v(p_1) 
 \e_\mu(k_1) \e_\n(k_2)\, ,\\
 {\cal M}_4 & = & -i {P_f \over u} \bar u(p_2)  [\slash{k}_1, \g^\m]  (\slash{k}_2 - \slash{p}_1)  \g^\n v(p_1) 
 \e_\mu(k_1) \e_\n(k_2)\, .
 \end{eqnarray}  
 For the $\xi = 1$ gauge we also need the matrix elements with the incoming FP-ghosts of the gluon. There are two graphs possible,
\begin{eqnarray}
 {\cal M}_{\eta} & = &  - i\, {P_f \over s}\, \bar u(p_2) \g^\r [\slash{k}_1 +  \slash{k}_2, \slash{k}_2] v(p_1) \, , \\ 
 {\cal M}_{\bar\eta} & = & - i\,  {P_f \over s}\, \bar u(p_2) \g^\r [\slash{k}_1 +  \slash{k}_2, \slash{k}_1] v(p_1)\, .
 \end{eqnarray} 
The result for the squared amplitudes for the massless case is
\begin{eqnarray}
|{\cal M}_{\eta}|^2 +  |{\cal M}_{\bar\eta}|^2  = -  4  {P_f \over s} C (s + 2 t)^2\, ,
\end{eqnarray}
with $C_3 = \sum_{a, b, c} |f^{a b c}|^2 = 24$ for $\suthree$.\\

For $\sq \sg \to \sq \grav$, named process H in table~\ref{table:1} we
have two graphs, with a $g$ propagator in the $t$-channel and 
a new one, with the four-point interaction squark-squark-gluino-goldstino, 
 \begin{eqnarray}
{\cal M}_{1} & = &  {P_f  \over t}  
\bar u(p_2) [\slash{k}_1 - \slash{p}_1, \slash{p}_1 + \slash{k}_1] u(k_2) \, , \\
{\cal M}_{2} & = &  2\, P_f\, \bar u(p_2) (P_R - P_L) u(k_2) \, ,
\end{eqnarray}   
with $P_R - P_L = \gamma_5$. Note that the quartic coupling is taken from~\cite{Pradler:2006tpx}, 
page~68, and  differs from 
this  in~\cite{Bolz:2000xi}, page~61.\\
The conjugated amplitudes read
 \begin{eqnarray}
{\cal M}^*_{1} & = & - {P_f  \over t}  
\bar u(k_2) [\slash{k}_1 - \slash{p}_1, \slash{p}_1 + \slash{k}_1] u(p_2) \, , \\
{\cal M}^*_{2} & = &  -2\, P_f\, \bar u(k_2) (P_R - P_L) u(p_2) \, ,
\end{eqnarray}
For the squared elements we get
\begin{eqnarray}
|{\cal M}_1|^2  & =  & - 8 {(2s + t)^2 \over t}\, \\
|{\cal M}_2|^2  & =  & - 8 t \, \\
{\cal M}_1 {\cal M}^*_2 &  = & - {\cal M}_2 {\cal M}^*_1 = 32 i \e_{\mu\nu\rho\d} k_1^\mu k_2^\nu p_1^\rho p_2^\d\, . 
\end{eqnarray}
The results read, including the prefactor and a factor of 2 for the two squark mass eigenstates,
\begin{eqnarray}
\label{eq:nondevH}
|{\cal M}_{\rm full}|^2 & = & - 4 C_3' \left( t + 2 s + {2 s^2 \over t} \right) {g_3^2 \over \mplanck^2} {M_{N_3}^2 \over 3 m^2_{3/2}}
\, ,\\
|{\cal M}_{\rm sub}|^2 & = & 0\, .
\end{eqnarray}

For $\sq_i g \to q \grav$, named process C in table~\ref{table:1} we
have only one graphs, with a $\sg$ propagator in the $t$-channel,
\begin{equation}
{\cal M}_{1}  =  i P_f {\sqrt{2} \over t} \bar u(p_1)  \a_{RL}^{i*}  (\slash{k}_1 - \slash{p}_1) [ \slash{k}_2, \g^\mu] v(p_2) \e_\mu(k_2)\, .
\end{equation}

For $q g \to q \grav$, named process G in table~\ref{table:1} we
have only one graphs, with a $g$~propagator in the $t$-channel,
\begin{equation}
{\cal M}_{1}  =  i {P_f  \over t}  
\bar u(p_1) \g_\mu u(k_1) \bar u(p_2) [ \slash{k}_1 - \slash{p}_1, \g^\mu] u(k_2)\, .
\end{equation}

\section{Thermal spectral functions}
\label{AppendixB}
For this calculation we have used the real time formalism (RTF)~\cite{Bellac:2011kqa,Kapusta:2006pm}, but we have also checked the validity of our results applying the imaginary time formalism. In the RTF the temperature dependent propagators read as
\bea
{\mathrm{ scalar:  }}\,\,\,\, && \D_S(K,T)= \frac{i}{K^2} + 2 \pi  \d(K^2)  n_{B}(K,T)\,,  \nn 
{\mathrm{ fermion:  }}\,\,\,\, && S_F(K,T) = \frac{i \, \slash{\!K}}{K^2} - \slash{\!K}\, 2 \pi  \d(K^2)  n_{F}(K,T)\,,    \nn
{\mathrm{ vector-boson : }}\,\,\,\, && \D^{ab}_{\m\n} (K,T)=  \d^{ab} g_{\m\n} 
        \left(  \frac{  - i }{K^2} - 2 \pi  \d(K^2)  n_{B}(K,T)      \right)\,, \nn
{\mathrm{ ghost  : }}\,\,\,\, && G^{ab} (K,T)=  \d^{ab}  
        \D_S(K,T),
 \label{eq:rtf_prop}       
\eea
where have assumed   massless particles,  at  the Feynman $\xi=1$ gauge.  In addition,  the Fermion/Boson particle densities are given by
\beq
n_{F/B}(K,T)=( e^{ |K_\m u^\m|/T} \pm 1)^{-1} \,,
\eeq
where the four-velocity at the plasma rest frame is $u^\m=(1,0,0,0)$.
\subsection{Vector-boson self-energy}
\label{Vector-boson self-energy}
The vector-boson self-energy consists of three different contributions, the scalar, the fermion and the vector-boson one.
\begin{figure}[h!]  
\centering{\includegraphics[bb=  151 625 450 770, scale=0.45]{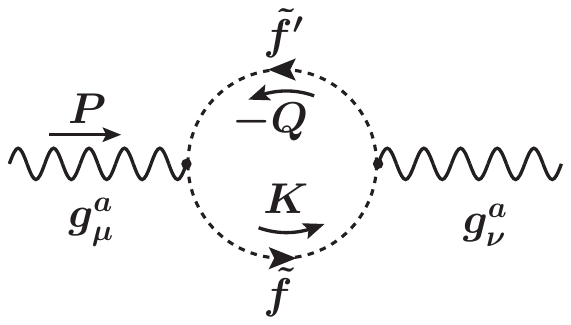} \quad \quad \includegraphics[bb=  151 625 450 770, scale=0.45]{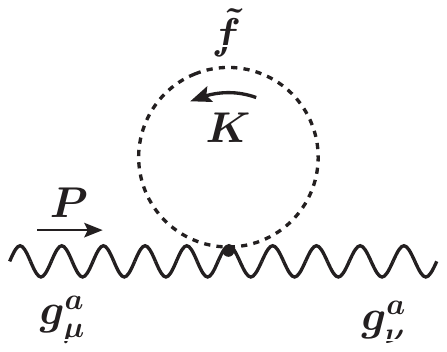} \quad \includegraphics[bb=  151 625 450 770, scale=0.45]{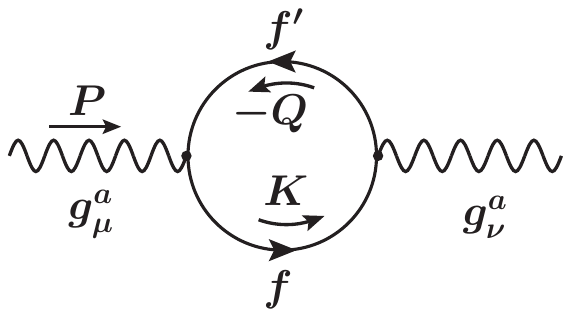}\vspace{8mm}
\includegraphics[bb=  151 625 450 770, scale=0.45]{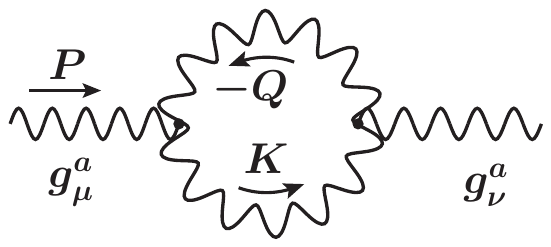} \quad \includegraphics[bb=  151 625 450 770, scale=0.45]{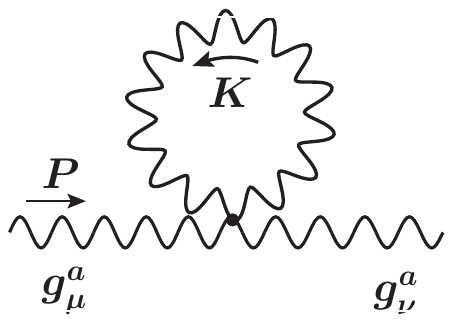} \quad
\includegraphics[bb=  151 625 450 770, scale=0.45]{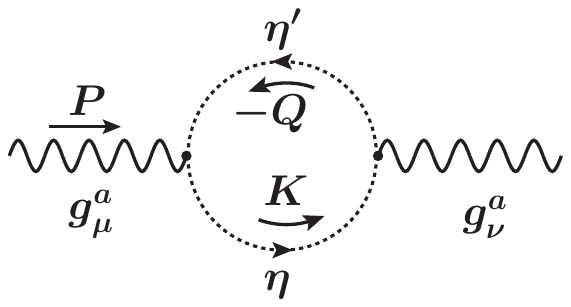}}
\caption{Feynman diagrams for the  vector-boson self-energy from the scalar contribution (\nth{1} - \nth{2}), the fermion contribution (\nth{3}) and vector-boson contribution (\nth{4} - \nth{6}). The momentum $Q$ is defined as $Q=P-K$.  
\label{fig:Vself}}
\end{figure}

 In figure~\ref{fig:Vself} we display these three different contributions, where the first two diagrams correspond to the scalar one, the third one is the unique diagram with a fermion in the loop and the rest three diagrams are those which are involved in the gauge-boson contribution. Below, we present the real parts of the self-energies $\P_{\m\n}$ which are needed in our calculation, where the superscripts S, F and V denote the  corresponding contributions, that is
 
 \begin{align}
\Rea[\Pi^{S}_{\mu\nu} (P)]  & = - g_N^2 N_S  \int {{\rm d}^4K \over  (2 \pi)^3} \; \left((2 K - P)_\mu (2 K - P)_\nu -(K-P)^2 g_{\mu\nu}\right) \nonumber
\\ 
 & \hspace{2cm}\times\frac{\delta(K^2) }{(K-P)^2}  n_B(K,T)\,, \nonumber
\\
\Rea [\Pi^{F}_{\m\n}(P)]  &=  - 4 g_N^2   N_F   \int \frac{{\rm d}^4 K}{(2\pi)^3} \left((K-P)_{(\m} K_{\n)}  + K_\r P^\r\,  g_{\m\n}\right) \frac{\d(K^2) }{(K-P)^2}   n_F(K,T)\,, \nonumber
\\
\Rea [\Pi^{V}_{\m\n}(P)]  &=   g_N^2\,  N_V \int \frac{{\rm d}^4 K}{(2\pi)^3} \left(  -4(K-P)_{(\m} K_{\n)} -  ( 2P^2  +4 K_\r P^\r ) \, g_{\m\n}  + 2P_\m P_\n\right)  \nonumber
\\
\hspace{2cm} & \times\frac{\delta(K^2)}{(K-P)^2 } n_B(K,T)\,. 
 \label{eq:pi_mu_mu}       
\end{align}
The numerical coefficients $N_S$, $N_F$ and $N_V$ for the MSSM are given in table~\ref{table:Ns}. Following the notation of~\cite{Rychkov:2007uq} we define the integrals
\begin{equation}
G_i \equiv  g^{\m\n}   \Rea (\Pi^i_{\m\n} )/ (g_N^2 \, N_i)\,, \qquad\text{and} \qquad
H_i \equiv  u^\m u^\n  \Rea (\Pi^i_{\m\n} )/ (g_N^2 \, N_i)
\end{equation}
where the subscript i stands for the three different contributions as discussed previously. After manipulations in agreement with~\cite{Rychkov:2007uq} we obtain that
\begin{align}
\label{eq:GandH}
G_S  = & \int_0^\infty {{\rm d}k \over 2 \pi^2} \left[ 2 k - {P^2 \over 8 p}  L_{-} \right]  n_B(K,T) = \frac{T^2}{6}  -\frac{P^2}{16\pi^2 p} I^B_1 \, ,\nn
G_F  = &    \int_0^\infty  {{\rm d }k \over 2 \pi^2}  \left[ 4 { k} +{P^2 \over {2{ p}}} L_{-}     \right]   n_F(K,T) = \frac{T^2}{6}  + \frac{P^2}{4\pi^2 p} I^F_1 \, , \nn
G_V  = &    \int_0^\infty  {{\rm d }k \over 2 \pi^2}  \left[ 4 { k} +{5\over 4 }{P^2 \over {{ p}}} L_{-}    \right]  n_B(K,T) = \frac{T^2}{3}  + \frac{5P^2}{8\pi^2 p} I^B_1\, , \\
H_S = &  \int_0^\infty {{\rm d}k \over 2 \pi^2} \left[  k \, L + {M \over 2p} + {p \over 8} L_{-} \right]  n_B(K,T) = \frac{T^2}{12} L  + \frac{p}{16\pi^2 } \left(1+\frac{P^2}{p^2}\right) I^B_1 
\nn
& \hspace{7.5cm} + \frac{1}{4\pi^2 p } I^B_2 +  \frac{p_0}{4\pi^2 p } I^B_3  \,, \nn
H_F   = & \int_0^\infty  {{\rm d} k \over 2 \pi^2}   \left[  2 { k}\, L  +{M  \over {{p}}}    \right]  n_F(K,T) = \frac{T^2}{12} L  + \frac{P^2}{8\pi^2 p}   I^F_1 
 + \frac{1}{2\pi^2 p } I^F_2 +  \frac{p_0}{2\pi^2 p } I^F_3 \, , \nn
H_V   = & \int_0^\infty  {{\rm d} k \over 2 \pi^2}   \left[  2 { k}\, L  +{M  \over {{p}}}  -{p\over 4} L_{-} \right]   n_B(K,T) =  \frac{T^2}{6} L   - \frac{p}{8\pi^2 } \left( 1 - \frac{P^2}{p^2}\right) I^B_1 
\nn
& \hspace{7.5cm} + \frac{1}{2\pi^2 p } I^B_2 +  \frac{p_0}{2\pi^2 p } I^B_3\, , \nonumber
\end{align}
with
\bea
L &=&1-\frac{p_0}{p}\ln \left(\frac{p_0+p}{p_0-p}\right)\,,\quad L_{\pm}(k)=\ln \left(\frac{2k+p_0+p}{2k+p_0-p}\right)\pm \ln \left(\frac{2k-p_0-p}{2k-p_0+p}\right)\,, \nn
% L_{\pm}(k)&=&\ln \left(\frac{2k+p_0+p}{2k+p_0-p}\right)\pm \ln \left(\frac{2k-p_0-p}{2k-p_0+p}\right)\,,\nn
M(k)& =& \left(\frac{p_0^2-p^2}{4}+k^2\right)L_{-}(k) + kp_0 L_{+}(k)\,.
\eea
The second equalities in~\eqref{eq:GandH} are obtained using the integrals

\beq
\label{eq:integrals}
\int_{0}^{\infty}\frac{{\rm d}k}{2\pi^2}k\, n_{B,F}(k)=\frac{T^2}{12}, \frac{T^2}{24} \quad \text{and} \quad I^{B,F}_{\{1,2,3\}} = \int_{0}^{\infty}{\rm d}k \{1,k^2,k \} L_{\{-,-,+\}}(k,p_0,p)\,n_{B,F}(k)\,.
\eeq
These  integrals $I_{1,2,3}^{F,B}$ have been used in our numerical analysis for the 
self-energies and the dispersion relations.

\subsection{Fermion self-energy}
\label{Fermion self-energy}
There are two Feynman diagrams (figure~\ref{fig:F_se}) contributing to the fermion selfenergy. One that is due to vector boson loop and another 
due to Yukawa type fermion loop. The relevant interaction Lagrangian is 
\beq
\label{eq:scferfer}
\mathcal{L}= -g_3 \, g^a_\m  \, \bar{f_s} \g^\m T^a_{st} \, f_t + \l_q \, \f\,\bar{f_s}  \G_{st}  f_t  \, .
\eeq
This yields the Feynman rules for the vertex  
 vector-boson($g^a_\mu$)-fermion($\bar{f_s}$)-fermion($f_t$)   $ = - i g_3 T^a_{st} \g_\mu$ and for the 
scalar($\phi$)-fermion($\bar{f_s}$)-fermion($f_t$) $ =  i \l \, \G_{st} $. 
%%%%%%%%
\begin{figure}[h!]  
\centering{\includegraphics[bb=   151 610 450 760, scale=0.53]{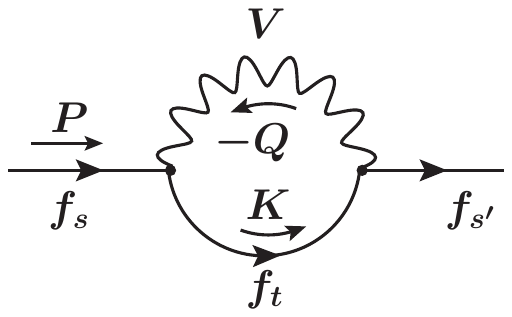} \, \includegraphics[bb=   151 610 450 760, scale=0.53]{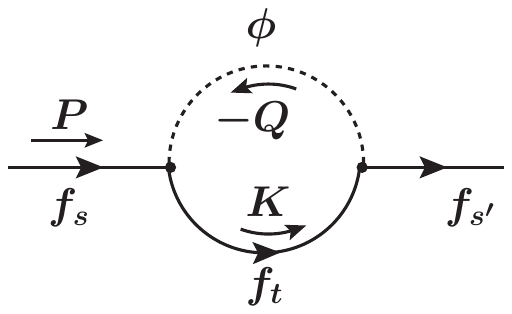}}
\caption{Feynman diagrams for the fermion self-energy from the vector-boson (left) and scalar (right) contribution. The momentum $Q$ is defined as $Q=P-K$.}
\label{fig:F_se}
\end{figure}
Using again the real time formalism propagators~\eqref{eq:rtf_prop} we obtain that the real part of the thermally corrected  fermion propagator (vector-boson part) is
  \beq
\Rea[{\S_V}(P)] = -  2  g_3^2  C_R \int \frac{ {\rm d}^4 K }{(2\pi)^3}
  \left[   \slash{\!K} \left(  n_B(K,T)  +n_F(K,T)   \right] - \slash{\!P}n_B(K,T)      \right) \frac{\d(K^2)}{(K-P)^2}  \,,
    \label{eq:imsv}
 \eeq
where $\S_a T^a_{s't} T^a_{ts}=C_R \d_{s's}$ is the Casimir operator in the representation $R$. The scalar part of the fermion self-energy  $\Rea[{\S_\f}]$, is the same as~\eqref{eq:imsv} if we substitute the overall factor $ 2  g_3^2 C_R \to   \lambda^2_q C' $, where $(\G \G^\dagger)_{s's}= C' \d_{s's}$~\cite{Weldon:1982bn}. 

For the $T=0$ part in each gauge group we have that\footnote{We have used that in $d=4-\epsilon$ dimensions $\frac{(2\pi^4)}{i \pi^2}\m^{4-d}\int {{\rm d}^d K \over (2 \pi)^d}  {\slashed{K}  \over K^2 (P - K)^2}  =  {i \over 16 \pi^2} {\slashed{P}  \over 2}  \left(\D - \ln{-P^2 \over \mu^2} + 2 \right)$ 
where $ \Delta =\frac{2}{\e}  -\g +\ln{4\pi} $ and $\gamma$ is the  Euler–Mascheroni constant. }

\begin{equation}
\Sigma_V^{(0)}(P) = - g_N^2 C_R \slashed{P} {1 \over 16 \pi^2} \left(2 - r - \ln{-P^2 \over \mu^2} -\g +\ln{4\pi} \right)\,,
\end{equation}
and similarly for $\Sigma_\f^{(0)}$.
Assuming the $\overline{\textrm{DR}}$ renormalization scheme ($r=0$) and consequently ignoring the terms $- \gamma + \ln 4\pi$, we obtain the spectral functions for fermion particles ($\rho_{+}$) and fermion holes ($\rho_{-}$) after some manipulations. These spectral functions, including the $T=0$ contribution, are given by equation~\eqref{eq:rhopm}.

\subsection{Dispersion relations}
\label{sec:disp_rel}
\subsubsection{Vector-bosons}
The $T\neq 0$ dispersion relation for longitudinal (transverse) modes are given by the poles of the longitudinal (transverse) parts of the vector propagator. For the longitudinal one we have to find the zeros of the equation $p^2(1-\pi_L/P^2)=0$ as it is dictated from the longitudinal spectral function~\eqref{eq:rhoLT}, that is
\bea
p^3&+& m_V^2 p\Big(1-\frac{p_0}{p}\ln\frac{p_0+p}{p_0-p}\Big) +\left(\frac{m_V}{T}\right)^2 \frac{3}{N_1 \pi^2} \Bigg[ {}  I_1^B \left( \frac{p_0^2}{4}N_2 - \frac{p^2}{2}N_c \right) \nn
& +& I_2^B N_2 +I_3^B p_0 N_2 +I_1^F \frac{p_0^2-p^2}{4}N_F  + I_2^F N_F +I_3^F p_0 N_F \Bigg]  =0\,,
\eea
where the integrals $I_{1,2,3}^{F,B}$ have been defined in~\eqref{eq:integrals} and the coefficients $N_1$ and $N_2$ are the following combinations
\beq
N_1= N_V+N_S/2+N_F/2 \qquad \text{and} \qquad N_2= N_V+N_S/2\,.
\eeq
Similarly\footnote{The HTL limits for the transverse and longitudinal modes are given by the equations $
 p^3+2m_V^2 p-m_V^2 p_0\ln\frac{p_0+p}{p_0-p} =0\,,$ and $p_0^2p^3-p^5 -m_V^2 p p_0^2 -m_V^2\frac{p_0}{2}(p^2-p_0^2)\ln\frac{p_0+p}{p_0-p} =0 $, respectively. These limits can be found in~\cite{Bellac:2011kqa,Kapusta:2006pm}.}, the dispersion relation for the transverse modes reads
\bea
p^3&-&m_V^2\frac{p}{2}\Big(1 - \frac{p_0}{p}\ln\frac{p_0+p}{p_0-p}\Big) +m_V^2\frac{p^3}{p^2-p_0^2} -\left(\frac{m_V}{T}\right)^2\frac{3}{2N_1\pi^2} \Bigg[  I_1^B \bigg( \frac{p_0^2-p^2}{4}N_2   \nn
 &+&p^2 N_c \bigg)+ I_2^B N_2 +I_3^B p_0 N_2 +I_1^F \frac{p_0^2+p^2}{4}N_F  + I_2^F N_F +I_3^F p_0 N_F \Bigg]  =0\,,
\eea 
 
\subsubsection{Fermions}
For the fermionic modes the $ T\neq 0 $  dispersion relations arise from the poles of the spectral functions~\eqref{eq:rhopm}, i.e. when $ (p_0 \mp p)/2 +\frac{m_F^2}{T^2}F_{\pm}=0 $. The $1-$loop dispersion relation for both $\rho_\pm$ takes the form\footnote{In the HTL approximation  the terms contain integrals can be dropped, so~\eqref{eq:polespm} reduces to $
 p_0 \mp p-\frac{m_F^2}{2p}\left[\left(1\mp \frac{p_0}{p}\right) \ln\frac{p_0+p}{p_0-p} \pm 2 \right]=0\,, $~\cite{Bellac:2011kqa,Kapusta:2006pm}.}
\bea
\label{eq:polespm}
 &&p_0 \mp p -\frac{m_F^2}{2p}\left[\left(1\mp \frac{p_0}{p}\right) \ln\frac{p_0+p}{p_0-p} \pm 2 \right]\nn
&& \mp \left(\frac{m_F}{T}\right)^2\frac{1}{\pi^2}\frac{p_0 \mp p}{p^2} \left[ I_3^B + \frac{p_0 \mp p}{2} I_1^B + I_3^F + \frac{p_0 \pm p}{2} I_1^F \right]=0\,.
\eea

\section{ Calculation of the collision term }
\label{AppendixC}
In this appendix we calculate the collision term for the 2$\rightarrow$2 interactions following closely the method applied in~\cite{Pradler:2006tpx}. Although, as we saw in section~\ref{2to2sub} only two different collision terms are needed for the calculation of the subtracted rate, we give a table with various results that may be useful in other interactions. All the numerical results shown in this appendix have been calculated and cross checked using various integration routines found in Cuba library~\cite{Hahn:2004fe}.

For the process $ a+b\rightarrow c +\Gr $ the collision term is given by
\begin{equation}
{\cal C} \equiv \gamma  =\frac{1}{(2\pi)^8} \int \frac{{\rm d^3} \mathbf{p_1}}{2E_1}  \frac{{\rm d^3} \mathbf{p_2}}{2E_2} \frac{{\rm d^3} \mathbf{p_3}}{2E_3} \frac{{\rm d^3} \mathbf{p}}{2E} \delta^4(P_1 + P_2 - P_3  - P) |{\cal M}|^2 f_a f_b (1 \pm f_c)\, .
\label{coll_term_general}
\end{equation}
Firstly, our aim is to calculate the quantity $  \frac{{{\rm d} \cal C}}{{\rm d^3} p}, $ where $ {\rm d^3} \mathbf{p} = {\rm d^3} p {\rm d}\Omega_p. $
Thus,
\begin{equation}\label{Pradler_int}
 \frac{{\rm d} {\cal C}}{{\rm d^3} p} =\frac{1}{2^9\pi^8 E} \int \frac{{\rm d^3} \mathbf{p_1}}{2E_1}  \frac{{\rm d^3} \mathbf{p_2}}{2E_2} \frac{{\rm d^3} \mathbf{p_3}}{2E_3} \int {\rm d}\Omega_p\, \delta^4(P_1 + P_2 - P_3  - P) |{\cal M}|^2 f_a f_b (1 \pm f_c)\, .
\end{equation}
This calculation will be done in the so-called $t$-frame, in which the reference momentum is the $t$-channel  momentum $ \mathbf{k}=\mathbf{p_1}-\mathbf{p_3} $. Of course the results are frame independent.
Thus we will express the other momenta defining first $\mathbf{\hat{k}} =\mathbf{\hat{z}}$,
\begin{equation}
\mathbf{k} =k(0,0,1)\,, \quad \mathbf{p} =E(0, \sin \tilde{\theta}, \cos \tilde{\theta})\,, \quad \mathbf{p}_{3} =E_{3}(\cos \phi \sin \theta, \sin \phi \sin \theta, \cos \theta)\,.
\end{equation}
In this frame the Mandelstam variables are
\begin{equation}
\begin{aligned} 
s &= (P_1+P_2)^2=(P_3+P)^2= 2E\,E_3 (1-\cos \theta \cos \tilde{\theta}-\sin \theta \sin \tilde{\theta} \sin \phi)\,,
\\ t &=(P_1-P_3)^2=(P-P_2)^2=(E_1-E_3)^2-k^2\,.
\end{aligned}
\end{equation}
Before we continue with the computation of~\eqref{Pradler_int}, we will prove some useful identities.
We will use the identity
\begin{equation}
\frac{{\rm d^3} \mathbf{p_1}}{2E_1}={\rm d^4}P_1 \,\delta^4(P_1^2)\, \Theta(E_1)={\rm d}E_1\,{\rm d^3}\mathbf{p_1} \,\delta^4(P_1^2)\, \Theta(E_1)\,,
\end{equation}
and we will insert $ \int {\rm d^3} \mathbf{k}\, \delta^3(\mathbf{k}-\mathbf{p_1}+\mathbf{p_3})=1, $ then
\begin{equation}
\frac{{\rm d^3} \mathbf{p_1}}{2E_1} = \int {\rm d^3} \mathbf{k}\, \delta^3(\mathbf{k}-\mathbf{p_1}+\mathbf{p_3}) {\rm d}E_1 {\rm d^3} \mathbf{p_1}\/ \delta^4(E_1^2-|\mathbf{p_1}^2|)\Theta(E_1)\,,
\end{equation}
and after integrating over $ {\rm d^3} \mathbf{p_1} $ using the $ \delta- $function we get
\begin{equation}\label{iddelta1t}
\frac{{\rm d^3}\mathbf{p_1}}{2E_1} =  \delta(E_1^2-|\mathbf{k}+\mathbf{p_3}|^2)\Theta(E_1)  {\rm d}E_1  {\rm d^3} \mathbf{k}\,.
\end{equation}
Another usefull identity is
\begin{equation} 
\frac{{\rm d^3} \mathbf{p_1}}{2E_1} =  {\rm d}E_1 {\rm d^3} \mathbf{p_1}\/ \delta^4(P_1^2) \Theta(E_1)
 = \frac{\delta(E_1-|\mathbf{p_1}|)}{2|\mathbf{p_1}|} \Theta(E_1) {\rm d}E_1 {\rm d^3} \mathbf{p_1}\, ,
\end{equation}
which after multiplying by $ \delta^4(P_1+P_2-P_3-P)  $ becomes
\begin{equation}\label{iddelta2t}
\frac{{\rm d^3} \mathbf{p_1}}{2E_1}\,\delta^4(P_1+P_2-P_3-P) = \delta((E_3+E-E_2)^2-|\mathbf{k}+\mathbf{p_3}|^2)\Theta(E_3+E-E_2)\,.
\end{equation}
In this calculation, we will use (\ref{iddelta1t}) as it is and (\ref{iddelta2t})  expressed in terms of $ \mathbf{p_2} $ and $ E_2 $
\begin{equation}\label{iddelta1p_3t}
\frac{{\rm d^3}\mathbf{p_2}}{2E_2}\,\delta^4(P_1+P_2-P_3-P) =  \delta((E_3+E-E_1)^2-|\mathbf{p}-\mathbf{k}|^2)\Theta(E_3+E-E_1)\,.
\end{equation}
Now, the next step is to rewrite the $ \delta- $functions in terms of $ \cos \theta  $ and $ \cos \tilde{\theta}. $ We have
\begin{align}
\label{eq:deltacosthetatildet}
\delta\left(E_{1}^{2}-|\mathbf{k}+\mathbf{p_3}|^{2}\right) &=\frac{1}{2 E_3 k} \delta\left(\cos \theta-\frac{E_1^{2}-E_{3}^{2}-k^{2}}{2 E_3 k}\right)\,, 
\\
\delta\left(\left(E+E_3-E_1\right)^{2}-\left|\mathbf{p}-\mathbf{k}\right|^{2}\right) &=\frac{1}{2  E k} \delta\left(\cos \tilde{\theta}-\frac{E^2+k^2-\left(E+E_3-E_1\right)^2}{2  E k}\right)\,. \nonumber
\end{align}

Substituting~\eqref{iddelta1t},~\eqref{iddelta1p_3t}, and~\eqref{eq:deltacosthetatildet} in~\eqref{Pradler_int} we get 
\begin{equation}\label{Pradler_int_againt}
\begin{aligned} 
 \frac{{\rm d} {\cal C}}{{\rm d^3} p}   = &\frac{1}{2^{12}\pi^8 E^2} \int {\rm d} \cos \tilde{\theta}\, {\rm d} \tilde{\phi}\,  {\rm d} \cos \theta\, {\rm d}  \phi \, {\rm d}E_1\, {\rm d} E_3 \,{\rm d} k\, {\rm d} \Omega_k
\\ & \times \delta\left(\cos \theta-\frac{E_1^{2}-E_{3}^{2}-k^{2}}{2 E_3 k}\right)\, \delta\left(\cos \tilde{\theta}-\frac{E^2+k^2-\left(E+E_3-E_1\right)^2}{2  E k}\right) 
\\ & \times |{\cal M}|^2 f_a f_b (1 \pm f_c) \Theta(E_1) \Theta(E_3) \Theta(E+E_3-E_1)\, .
\end{aligned} 
\end{equation}
Nothing depends on $ {\rm d} \tilde{\phi} $ and $ {\rm d} \Omega_k, $ so after these integrations we get an additional $ 8\pi^2 $ factor. 
After the $\theta$ and $\tilde{\theta}$ integrations  we have to substitute 
\begin{equation} 
 \cos \theta=\frac{E_1^{2}-E_{3}^{2}-k^{2}}{2 E_3 k}\qquad \text{and} \qquad \cos \tilde{\theta}=\frac{E^2+k^2-\left(E+E_3-E_1\right)^2}{2  E k}\,.  \\ 
\label{angle0t}
\end{equation} 
From the integrations over the $ \delta- $functions, we find that
\begin{equation}
-1\leq \cos \theta \leq 1 \Rightarrow 
     \begin{cases}
      E_3-E_1\leq k \leq E_1+E_3\\
       E_1-E_3\leq k\,,
     \end{cases}
\end{equation}
\begin{equation}
\hspace{10mm}-1\leq \cos \tilde{\theta} \leq 1 \Rightarrow 
     \begin{cases}
      E_1-E_3\leq k \leq 2E+E_3-E_1\\
       E_3-E_1\leq k\,,
     \end{cases}
\end{equation}
which yield the $ \Theta- $functions $ \Theta(k-|E_1-E_3|), $  $ \Theta(E_1+E_3-k) $ and $ \Theta(2E+E_3-E_1-k). $\\
After performing the remaining angular integrations we find that
\begin{equation}\label{Pradler_int_finalt}
\frac{d {\cal C}}{{\rm d^3} p} =\frac{1}{2^9\pi^6 E^2} \int  {\rm d}E_1\, {\rm d} E_3 \, {\rm d} q \,{\rm d}  \phi \,
|{\cal M}|^2 f_a f_b (1 \pm f_c) {\vartheta}\,,
\end{equation}
where $\vartheta$ is defined as

\begin{equation}\label{Thetaft}
\begin{aligned}
\vartheta  =  &\Theta(k-|E_1-E_3|) \Theta(E_1+E_3-k) 
\\ & \times\Theta(2E+E_3-E_1-k)\Theta(E+E_3-E_1)\Theta(E_1) \Theta(E_3)\,,
\end{aligned} 
\end{equation}
and in $|{\cal M}|^2 $ we have substituted $\theta$ and $\tilde{\theta}$ from~(\ref{angle0t}).
Using the identities for the Heaviside step function 
\begin{align}
\Theta(E_1+E_3-k)=& 1-\Theta(k-E_1-E_3)\,,
\\
\Theta(k-E_1-E_3)\Theta(k-|E_1-E_3|)=& \Theta(k-E_1-E_3)\,,
\end{align}
and substituting $1=\Theta(E_1-E_3)+\Theta(E_3-E_1)$ in equation~\eqref{Thetaft}, we obtain
\begin{equation}\label{Thetaf3t}
\begin{aligned}
{\vartheta}  = {} & \Theta(k-E_1+E_3)\Theta(2E+E_3-E_1-k)\Theta(E+E_3-E_1)\Theta(E_1-E_3) \Theta(E_1) \Theta(E_3)
\\ & +\Theta(k+E_1-E_3)\Theta(2E+E_3-E_1-k)\Theta(E+E_3-E_1)\Theta(E_3-E_1) \Theta(E_1) \Theta(E_3)
\\ & - \Theta(k-E_1-E_3)\Theta(2E+E_3-E_1-k)\Theta(E+E_3-E_1)\Theta(E_1) \Theta(E_3)\, .
\end{aligned} 
\end{equation}
Note that $\Theta(k-E_1-E_3)\Theta(2E+E_3-E_1-k) \Rightarrow E_{1}<E$, so we have to include the corresponding $ \Theta- $function. It is obvious that we will need to calculate 3 different integrals because of the 3 different combinations 
 of $ \Theta- $functions in~(\ref{Thetaf3t}). We have 
\begin{equation}\label{Thetaf4t}
\begin{aligned}
{\vartheta}  ={\vartheta}_{1}+{\vartheta}_{2}+{\vartheta}_{3} \,, 
\end{aligned} 
\end{equation}
where
\begin{equation}\label{Thetax4t}
\begin{aligned}
&{\vartheta}_{1} =  \Theta(k-E_1+E_3)\Theta(2E+E_3-E_1-k)\Theta(E+E_3-E_1)\Theta(E_1-E_3) \Theta(E_1) \Theta(E_3)\,,   \\
&{\vartheta}_{2} =  \Theta(k+E_1-E_3)\Theta(2E+E_3-E_1-k)\Theta(E+E_3-E_1)\Theta(E_3-E_1) \Theta(E_1) \Theta(E_3)\,, \\
&{\vartheta}_{3}= -  \Theta(k-E_1-E_3)\Theta(2E+E_3-E_1-k)\Theta(E+E_3-E_1)\Theta(E-E_1)\Theta(E_1) \Theta(E_3) \,.
\end{aligned} 
\end{equation}
After splitting the ${\vartheta}$ into three terms as in~(\ref{Thetaf4t})
we can write
\begin{table}[t!]
\centering
\begin{tabular}{c c }
\hline \hline
     \rowcolor{gray!15} 
$A$ & $k$-limits   \\
\hline  \\[-3.5mm]
1 & $ E_1-E_3 \le k \le 2E+E_3-E_1 $ \\
2 & $ E_3-E_1 \le k \le 2E+E_3-E_1 $ \\ 
3 & $ E_1+E_3 \le k \le 2E+E_3-E_1 $ \\
\hline
\end{tabular}
\caption{The integration limits for $k$ for the various cases.}
\label{table:klimits}
\end{table}
\begin{equation}
\frac{{\rm d} {\cal C}^{|{\cal M}|^2 }}{{\rm d^3} p} = \sum_A  g_{A}^{|{\cal M}|^2} \, ,
\end{equation}
where $|{\cal M}|^2=\{s,t,s^2,t^2,s t \}$ and $A=\{1,2,3 \}$.
We have defined 
\begin{equation}
  g_{A}^{|{\cal M}|^2} =    \frac{1}{2^9\pi^6 E^2}   \int_0^{\infty} {\rm d}E_3 \int_{0}^{E+E_3}  {\rm d}E_1 \int {\rm d} k \, 
      \int_{0}^{2\pi}  {\rm d}\phi \,  |{\cal M}|^2  \,  f_{abc}  \,{\vartheta}_{A}  \, ,
\label{gfunt}
\end{equation}
\begin{table}[t!]
\centering
\begin{tabular}{cccc}
\hline \hline
     \rowcolor{gray!15} 
$|{\cal M}|^2 $& ${\rm BBF} $& ${\rm BFB}$ & ${\rm FFF}$  \\
\hline  \\[-3.2mm]
${s}$ & $0.260\cdot 10^{-3}$ & $0.271 \cdot 10^{-3}$&  $0.151 \cdot 10^{-3} $  \\
${t}$ & $-0.130 \cdot 10^{-3}$ & $-0.133 \cdot 10^{-3}$ & $-0.756 \cdot 10^{-4}$ \\
${s^2}$ &$ 0.540 \cdot 10^{-2}$ & $0.546 \cdot 10^{-2}$ & $0.418 \cdot 10^{-2} $   \\
${t^2}$ & $0.180 \cdot 10^{-2}$ & $0.181 \cdot 10^{-2}$ & $0.140 \cdot 10^{-2}$   \\
${st}$ & $-0.270 \cdot 10^{-2}$ & $-0.271 \cdot 10^{-2}$ & $-0.209 \cdot 10^{-2}$  \\
 \hline 
\end{tabular}
\caption{The values of the collision  term  normalised by $1/T^6$ (for $s$ and $t$) or $1/T^8$ (for $s^2$, $t^2$ and $st$), for the possible statistical factors and the five basic squared amplitudes.}
\label{tbl_camp}
\end{table}
and the statistical factor is $f_{abc}  =   f_a(E_1) f_b(E_2)  (1 \pm f_c(E_3))$.
% \begin{equation}
% f_{abc}  =   f_a(E_1) f_b(E_2)  (1 \pm f_c(E_3))  \, .
% \end{equation}
The integration limits for the ${\rm d} k  $ integration are dictated by the theta functions in~(\ref{Thetax4t}).
As we have mentioned 
we will calculate 
five different amplitudes  $ |{\cal M}|^2 $ for completeness. These are $ |{\cal M}|^2=s,t,s^2,t^2 $ and $ st$.
Moreover, since we will integrate analytically over $k$ and $\phi$, it will be useful to define the function 
\begin{equation}
  \tilde{g}_{A}^{|{\cal M}|^2}(E,E_1,E_3) =  \frac{1}{2\pi}  \int {\rm d} k \,     \int_{0}^{2\pi}  {\rm d}\phi \,  |{\cal M}|^2  \, .
  \label{gtilt}
\end{equation}
Thus from~(\ref{gfunt})  and~(\ref{gtilt}),  we obtain that
\begin{equation}
  g_{A}^{|{\cal M}|^2} =    \frac{1}{2^8\pi^5 E^2}   \int_0^{\infty} {\rm d}E_3 \int_{0}^{E+E_3}  {\rm d}E_1 \,   \tilde{g}_{A}^{|{\cal M}|^2}(E,E_1,E_3)  \,  f_{abc}  \,{\vartheta}_{A}  \, .
\end{equation}
Using now the limits from the table~\ref{table:klimits} based on~(\ref{Thetax4t}), we are ready to calculate  $ \tilde{g}_{A}^{|{\cal M}|^2}(E,E_1,E_3)$. In details we have\\

$ \bullet $ For $ |{\cal M}|^2 =s $  
\begin{equation}
\begin{aligned}
& \tilde{g}_{1}^s  =  \frac{4}{3} \left(E-E_1+E_3\right)^2\left(E+2E_1+E_3\right)\,,  \\
& \tilde{g}_{2}^s   =  \frac{4}{3} E^2 \left(E+3 E_3\right)\,,  \\
& \tilde{g}_{3}^s  = \frac{4}{3} (E-E_1)\left(E^2+3E_3(E+E_1)+E E_1-2E_1^2\right)\,. 
\end{aligned}
 \end{equation}

$ \bullet $ For $ |{\cal M}|^2 =t $  
\begin{equation}
\begin{aligned}
& \tilde{g}_{1}^t  =  -\frac{4}{3} \left(E-E_1+E_3\right)^2\left(2E+E_1-E_3\right)\,,  \\
 &\tilde{g}_{2}^t   = -\frac{4}{3} E^2 \left(2E-3E_1+3E_3\right)\,,  \\
 &\tilde{g}_{3}^t  = -\frac{4}{3} (E-E_1) \left(3E_3(E+E_1)+(E-E_1)(2E+E_1)\right)\,. 
\end{aligned}
\end{equation}

$ \bullet $ For $ |{\cal M}|^2 =s^2 $  
\begin{align}
& \tilde{g}_{1}^{s^2}  = \frac{16}{15} \left(E-E_1+E_3 \right)^3\left(E^2+3E E_1 +2E E_3 +6E_1^2+3E_1 E_3 + E_3^2  \right)\,, \nonumber  \\
& \tilde{g}_{2}^{s^2}   = \frac{16}{15} E^3 \left(E^2+5 E E_3+10 E_3^2\right)\,,  \\
& \tilde{g}_{3}^{s^2}   =\frac{16}{15} \left(5E_3(E^4-4E E_1^3 +3E_1^4)+10E_3^2(E^3-E_1^3)+(E^2+3E E_1+6E_1^2)(E-E_1)^3\right)\,.\nonumber
\end{align}

\begin{table}[t!]
\centering
\begin{tabular}{cccc}
\hline \hline
     \rowcolor{gray!15} 
$|{\cal M}|^2$& ${\rm BBF} $& ${\rm BFB}$ & ${\rm FFF}$  \\
\hline  \\[-3.2mm]
${s}$ & $0.295\cdot 10^{-3} \,(+14\%)$ & $0.221  \cdot 10^{-3} \,(-18 \%)$&  $0.166 \cdot 10^{-3}\, (+10 \%) $ \\
${t}$ & $-0.148  \cdot 10^{-3}\,(+14 \%)$ &$ -0.111 \cdot 10^{-3}\, (-17 \%)$ & $-0.830 \cdot 10^{-4}\, (+10 \%)$ \\
${s^2}$ & $0.574 \cdot 10^{-2}\,(+6\%) $& $0.502 \cdot 10^{-2} \,(-8 \%)$ & $0.440 \cdot 10^{-2}\, (+5 \%)$   \\
${t^2}$ & $0.191 \cdot 10^{-2} \,(+6 \%) $& $0.167 \cdot 10^{-2}\,(-7 \%)$ & $0.147 \cdot 10^{-2} \,(+5 \%)$  \\
${st}$ & $-0.287 \cdot 10^{-2}\, (+6 \%) $& $-0.251 \cdot 10^{-2} \,(-7 \%)$ & $-0.220 \cdot 10^{-2}\, (+5 \%)$  \\
 \hline 
\end{tabular}
\caption{The $\cal{C}'$  collision term normalised by $1/T^6$
 (for $s$ and $t$) or $1/T^8$ (for $s^2$, $t^2$ and $st$), where the statistical factor $f_c$ has been neglected. 
The percentages in the parentheses are the deviations from the value of $\cal{C}$, namely  $ (\cal{C}'-\cal{C})/\cal{C} \,  \% $.}
\label{tbl_campp}
\end{table}

$ \bullet $ For $ |{\cal M}|^2 =t^2 $  
\begin{align}
&\tilde{g}_{1}^{t^2}  =\frac{16}{15} \left( 6E^2+3E(E_1-E_3)+(E_1-E_3)^2\right)\left( E-E_1+E_3\right)^3\,,  \nonumber
\\
&\tilde{g}_{2}^{t^2}   = \frac{16}{15} E^3 \left(6E^2+15E(E_3-E_1)+10(E_1-E_3)^2\right)\,,  
\\
&\tilde{g}_{3}^{t^2}   =\frac{16}{15} \left(6E^5+15E^4(E_3-E_1)+10E^3(E_1-E_3)^2-E_1^3(E_1^2-5E_1 E_3+10E_3^2)\right)\,. \nonumber
 \end{align}

$ \bullet $ Finally, for $ |{\cal M}|^2 =s\, t $  
% \begin{equation}
\begin{align}
& \tilde{g}_{1}^{s\, t}  =-\frac{16}{15} (E-E_1+E_3)^3 \left(3 E^2+E (4
   E_1+E_3)+(E_1-E_3) (3 E_1+2 E_3)\right)\,, \nonumber
   \\
& \tilde{g}_{2}^{s\, t}   = -\frac{16}{15} E^3 \left(3 E^2-5 E (E_1-2 E_3)+10
   E_3 (E_3-E_1)\right)\,, 
   \\
& \tilde{g}_{3}^{s\, t}   =-\frac{16}{15} \Big[ 10 E_3^2 \left(E^3-E_1^3\right)+10 E_3
   \left(E^2+E E_1+E_1^2\right) (E-E_1)^2  \nonumber \\
   &  + \left(3 E^2+4 E E_1+3 E_1^2\right) (E-E_1)^3 \Big] \, . \nonumber
\end{align}
 % \end{equation}

Based on the previous analytical results for the $ \tilde{g}_{A}^{|{\cal M}|^2}(E,E_1,E_3)$ we will use the relation 
\begin{equation}
 \mathcal{C}_{abc}^{|{\cal M}|^2} =   
  \frac{1}{2^8\pi^5}  \int_0^{\infty} {\rm d}E  \int_0^{\infty} {\rm d}E_3 \int_{0}^{E+E_3}  {\rm d}E_1 \, 
   \sum_A \{ \tilde{g}_{A}^{|{\cal M}|^2}(E,E_1,E_3)  \,{\vartheta}_{A}  \}  \,  f_{abc}  \, ,
\end{equation}
in order to perform numerically the integrations over $E_1,E_3$ and $E$. The ${\vartheta}_A$ is taken from~(\ref{Thetaf4t}) and the statistical 
factor $f_{abc}$ can be $f_{BBF}$,  $f_{BFB}$ or  $f_{FFF}$. 
For all these cases  the numerical values for the collision terms, normalised by $1/T^6$ or $1/T^8$
are summarized in the table~\ref{tbl_camp}.
For our calculation of the subtracted rate we need only the numerical factors $ {\cal C}_{\scriptscriptstyle \rm BBF}^s = 0.25957 \times 10^{-3} $ and 
${\cal C}_{\scriptscriptstyle \rm BFB}^t = -0.13286 \times 10^{-3}$.

Moreover ignoring the statistical factor for the accompanying particle of the gravitino, that is $ 1 \pm f_c(E_3) =1$, we  can calculate the collision factor $\cal{C}'$ defined as
\begin{equation}
{\cal C}' =\frac{1}{(2\pi)^8} \int \frac{{\rm d^3} \mathbf{p_1}}{2E_1}  \frac{{\rm d^3} \mathbf{p_2}}{2E_2} \frac{{\rm d^3} \mathbf{p_3}}{2E_3} \frac{{\rm d^3} \mathbf{p}}{2E} \delta^4(P_1 + P_2 - P_3  - P) |{\cal M}|^2 f_a f_b\,,
\end{equation}
analytically. The numerical values for ${\cal C}'$ along with the deviations from the value of ${\cal C}$ are given in table~\ref{tbl_campp}.

%  \vspace{-0.5 cm}
\bibliography{gravitino_refs}
\end{document}